\documentclass[reprint, amsmath,amssymb,aps,prb,superscriptaddress]{revtex4-1}
\usepackage[dvipdfmx]{graphicx}
\usepackage{array,multirow}
\usepackage{graphicx}
\usepackage{dcolumn}
\usepackage{bm}
\usepackage{color}

\begin{document}

\preprint{}
\title{Ground state properties of $\mathrm{Na_2IrO_3}$ \\
determined from {\it
ab initio} Hamiltonian  
and its extensions \\  
containing Kitaev and extended Heisenberg interactions}
\author{Tsuyoshi Okubo}
 \email{t-okubo@issp.u-tokyo.ac.jp}
\affiliation{Institute for Solid State Physics, University of Tokyo,
 Kashiwa, Chiba 277-8581, Japan
}%
\author{Kazuya Shinjo}%
\affiliation{Yukawa Institute for Theoretical Physics, Kyoto University, Kyoto
 606-8502, Japan}%
\affiliation{Computational Condensed Matter Physics Laboratory, RIKEN, Saitama 351-0198, Japan}%
\author{Youhei Yamaji}%
\affiliation{Quantum-Phase Electronics Center (QPEC), The University of Tokyo,
Tokyo, 113-8656, Japan}
\author{Naoki Kawashima}%
\affiliation{Institute for Solid State Physics, University of Tokyo,
 Kashiwa, Chiba 277-8581, Japan
}%
\author{Shigetoshi Sota}%
\affiliation{Computational Materials Science Research Team,
RIKEN Advanced Institute for Computational Science (AICS), Kobe, Hyogo 650-0047, Japan}%
\author{Takami Tohyama}%
\affiliation{Department of Applied Physics, Tokyo University of Science,
 , Tokyo 125-8585, Japan
}%
\author{Masatoshi Imada}%
\affiliation{Department of Applied Physics, University of Tokyo, Tokyo 113-8656,
 Japan}%
\date{\today}

\begin{abstract}
 We investigate the ground state properties of $\mathrm{Na_2IrO_3}$
 based on numerical calculations of the recently proposed {\it ab
 initio} Hamiltonian represented by Kitaev and extended Heisenberg
 interactions. To overcome the limitation posed by small tractable
 system sizes in the exact diagonalization study employed in a previous
 study (Yamaji {\it et al.}, Phys. Rev. Lett. {\bf 113}, 107201 (2014)),
 we apply two-dimensional density matrix renomalization group, and
 infinite-size tensor-network method.  By calculating at much larger
 system sizes, we critically test the validity of the exact
 diagonalization results.  The results consistently indicate that the
 ground state of $\mathrm{Na_2IrO_3}$ is a magnetically ordered state
 with zigzag configuration in agreement with experimental observations
 and the previous diagonalization study.  Applications of the two
 independent methods in addition to the exact diagonalization study
 further uncover a consistent and rich phase diagram near the zigzag
 phase beyond the accessibility of the exact diagonalization.  For
 example, in the parameter space away from the {\it ab initio} value of
 $\mathrm{Na_2IrO_3}$ controlled by the trigonal distortion, we find
 three phases: (i) an ordered phase with the magnetic moment aligned
 mutually in $120$ degrees orientation on every third hexagon, (ii) a
 magnetically ordered phase with a $16$-site unit-cell, and (iii) an
 ordered phase with presumably incommensurate periodicity of the moment.
It suggests that potentially rich magnetic structures may appear in
 $A_\mathrm{2}\mathrm{IrO_3}$ compounds for $A$ other than Na. The
 present results also serve to establish the accuracy of the
 first-principles approach in reproducing the available experimental
 results thereby further contribute to find a route to realize the
 Kitaev spin liquid.
\end{abstract}

\pacs{}
\maketitle

\section{Introduction\label{Sec:Intro}}

Novel quantum phenomena induced by strong spin-orbit interaction have
recently attracted much interest in condensed matter physics. Iridium
oxides offer a typical example that shows rich and interesting phenomena
\cite{JackeliK2009,ChaloupkaJK2010, WanTVS2011,WitczakCKB2014}. Among
them, $A_\mathrm{2}\mathrm{IrO_3}$ ($A =$ Na or Li) have most
intensively been investigated
\cite{Liu2011,Choi2012,Ye2012,LoveseyD2012,AlpichshevMCG2015,SinghMRBTKTG2012}
since the theoretical proposal that the Kitaev spin liquid would be realized~\cite{JackeliK2009,ChaloupkaJK2010}. 

The Kitaev interaction is an anisotropic Ising-like interaction, $S^\gamma
S^\gamma$, with the easy axes $\gamma$ depending on the direction of the
interacting bonds. For the model represented only by the Kitaev interaction called Kitaev model, the ground state is proved to be a quantum
spin-liquid \cite{Kitaev2006}.  As a more realistic model
representing $\mathrm{Na_2IrO_3}$, the so-called Kitaev-Heisenberg model
with both the Heisenberg and Kitaev interactions has been
proposed~\cite{JackeliK2009,ChaloupkaJK2010}. However, it has turned out that this model
does not either properly account for the experimental observation of the zigzag magnetic
order stabilized at low temperatures
\cite{Choi2012,Ye2012,LoveseyD2012}.  In order to bridge the discrepancy between
the experiments and the theoretical predictions, several alternative models have been
proposed. They contain further neighbor interactions
\cite{Choi2012,Ye2012,KimuchiY2011,ChaloupkaJK2013,SizyukPWP2014} or
additional anisotropic interactions caused by the trigonal distortion
\cite{BhattacharjeeLK2012,RauLK2014,KatukuriNYSK2014,YamajiNKAI2014,RauK2014arXiv}.

In this paper, in order to further clarify the nature of
$\mathrm{Na_2IrO_3}$, we investigate the ground state of the {\it ab
initio} Hamiltonian for $\mathrm{Na_2IrO_3}$ proposed by Yamaji {\it et
al.}  \cite{YamajiNKAI2014} and summarized in Appendix
\ref{App:Hamiltonian}. In the {\it ab initio} Hamiltonian, where
off-diagonal anisotropic interactions due to the trigonal distortion as
well as weak second-nearest neighbor and third-nearest neighbor
interactions are nonzero beyond the simple Kitaev-Heisenberg
Hamiltonian, the experimentally observed zigzag order was reproduced by
exact diagonalizations (ED) of clusters of $24$ and $32$
sites~\cite{YamajiNKAI2014,YamajiSYSKI2016}. However, more thorough studies are desired
beyond small clusters to understand intrinsic properties in the
thermodynamic limit.

In the present article,
we carry out larger size calculations using sophisticated numerical
methods; density matrix renomalization group (DMRG) and tensor network
(TN). 
Applicability of the newly developed TN method to
 {\it ab initio} Hamiltonians containing complex interactions beyond simple model
Hamiltonians\cite{IreguiCT2014}, is examined by carefully comparing with the ED and DMRG.  If the trigonal distortion is close to that of the {\it ab initio} Hamiltonian for  $\mathrm{Na_2IrO_3}$, the two methods show the zigzag order consistently with the exact diagonalization results in Ref.\onlinecite{YamajiNKAI2014}. 

Another purpose of this study is to clarify the role of the trigonal distortion and search phases competing with the zigzag order when the trigonal distortion is deviated from that of $\mathrm{Na_2IrO_3}$.
 Richer phase diagram beyond the exact diagonalization is determined,
where several distinct symmetry broken magnetic orders emerge, when the trigonal distortion is monitored away from $\mathrm{Na_2IrO_3}$ (as introduced in Eqs.~\eqref{eq:Hamiltonian} and \eqref{eq:trigonal}, later). This study may have relevance to other materials
$A_\mathrm{2}\mathrm{IrO_3}$ for $A$ other than Na because the trigonal distortion depends on $A$.

The rest of the paper is organized as follows. In Sec.~\ref{Sec:Model},
we describe our model. In Sec. \ref{Sec:Classical}, we present 
results of classical approximations. Our main results for the {\it ab initio} Hamiltonian and its derivatives are presented in Sec.~\ref{Sec:GroundState}. Finally, we summarize
our results in Sec.~\ref{Sec:Conclusions}.
\section{Model\label{Sec:Model}}
\begin{figure}
  \includegraphics[width=8cm]{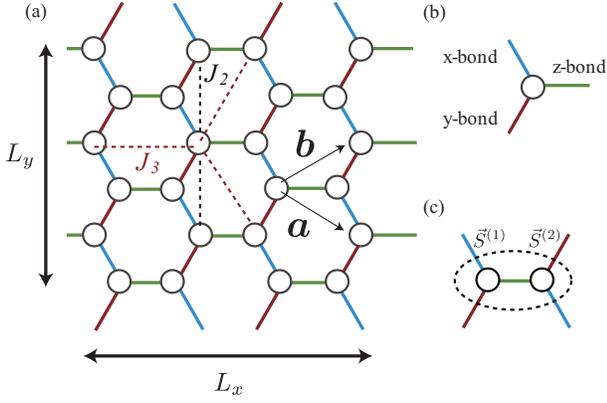}
  \caption{(a) Two-dimensional honeycomb lattice where $J_2$ and $J_3$
  represent the second- and the third-neighbor interaction pairs. In
  numerical calculations, we mainly use $L_x \times L_y$ lattices or
  unit cells with periodic boundary condition for both of $L_x$ and
  $L_y$ directions. (b) Three directions of the nearest-neighbor
  interactions. (c) Definition of the unit used in the classical analysis.\label{fig:unit_cells}}  
\end{figure}

The Hamiltonian of the model we investigate is given by
\begin{equation}
\mathcal{H} \equiv  \mathcal{H}_{\mathrm{1st}} +
 \mathcal{H}_{\mathrm{2nd}} + \mathcal{H}_{\mathrm{3rd}},
 \label{eq:Hamiltonian_spin}
\end{equation}
where $\mathcal{H}_{\mathrm{1st}}$, $\mathcal{H}_{\mathrm{2nd}}$, and 
$\mathcal{H}_{\mathrm{3rd}}$ express the nearest neighbor, the second
neighbor, and the third neighbor interactions on the honeycomb lattice,
respectively (see Fig.~\ref{fig:unit_cells}). For the
nearest neighbor interaction, we consider distorted Kitaev-Heisenberg
interaction with off-diagonal terms as 
\begin{equation}
 \mathcal{H}_{1st} \equiv \sum_{\Gamma = X,Y,Z}\sum_{\langle i,j\rangle\in \Gamma} \vec{S}_i^{\mathrm{T}} J_{\Gamma} \vec{S}_j, 
\end{equation}
where $\Gamma$ means the directions of the interactions and real
symmetric matrices $J_{\Gamma}$ ($\Gamma = X,Y,Z$) are give by
\begin{align}
 J_{X} & \equiv \begin{pmatrix}
		 K'&I_2''&I_2'\\
		 I_2''&J''&I_1'\\
		 I_2'&I_1'&J'
		\end{pmatrix},&
 J_{Y} & \equiv \begin{pmatrix}
		 J''&I_2''&I_1'\\
		 I_2''&K'&I_2'\\
		 I_1'&I_2'&J'
		\end{pmatrix},\notag\\
 J_{Z} & \equiv \begin{pmatrix}
		 J&I_1&I_2\\
		 I_1&J&I_2\\
		 I_2&I_2&K
		\end{pmatrix}.
\end{align}
For the second neighbor interactions, we consider only the interaction
perpendicular to the $Z$ bond:
\begin{equation}
\mathcal{H}_{\mathrm{2nd}} \equiv \sum_{\langle\langle i,j\rangle\rangle\in Z'} \vec{S}_i^{\mathrm{T}} J_{Z'}^{(\mathrm{2nd})} \vec{S}_j,
\end{equation}
where $\sum_{\langle\langle i,j\rangle\rangle\in Z'}$ represents the sum over
second neighbor pairs perpendicular to the $Z$ bond, and
$J_{Z'}^{(2nd)}$ is give by
\begin{equation}
  J_{Z'}^{\mathrm(2nd)}  \equiv J^{\mathrm(2nd)} = \begin{pmatrix}
		 J^{\mathrm(2nd)}&I_1^{\mathrm(2nd)}&I_2^{\mathrm(2nd)}\\
		 I_1^{\mathrm(2nd)}&J^{\mathrm(2nd)}&I_2^{\mathrm(2nd)}\\
		 I_2^{\mathrm(2nd)}&I_2^{\mathrm(2nd)}&K^{\mathrm(2nd)}
		\end{pmatrix}.
\end{equation}
Finally, for the third neighbor interaction the Hamiltonian is given by
\begin{equation}
 \mathcal{H}_{\mathrm{3rd}} \equiv \sum_{\Gamma = X,Y,Z}\sum_{\langle\langle\langle i,j\rangle\rangle\rangle\in \Gamma} \vec{S}_i^{\mathrm{T}} J_{\Gamma} \vec{S}_j, 
\end{equation}
where $\sum_{\langle\langle\langle i,j\rangle\rangle\rangle\in \Gamma}$
 represents the sum over the third neighbor pairs parallel to the $\Gamma$
 direction ($\Gamma = X,Y,Z$). We consider only the isotropic Heisenberg
 interaction for the third neighbor interaction:
\begin{equation}
 J_{\Gamma}^{\mathrm(3rd)}  \equiv J^{\mathrm(3rd)} = \begin{pmatrix}
				    J^{\mathrm{(3rd)}}&0&\\
				    0&J^{\mathrm{(3rd)}}&0\\
				    0&0&J^{\mathrm{(3rd)}}
				   \end{pmatrix}.
\end{equation}

The above spin Hamiltonian has been derived from the {\it ab initio}
Hamiltonian for $t_{2g}$ electrons of iridium atoms in
$\mathrm{Na_2IrO_3}$:
\begin{equation}
 \hat{H}_{t2g} = \hat{H}_0 + \hat{H}_{\mathrm{tri}} +
  \hat{H}_{\mathrm{SOC}} + \hat{H}_{U},
  \label{eq:Hamiltonian}
\end{equation}
where $\hat{H}_0$, $\hat{H}_{\mathrm{tri}}$, $\hat{H}_{\mathrm{SOC}}$
and $\hat{H}_{U}$ represent the hopping term, the trigonal distortion
with orbital-dependent chemical potentials, the spin-orbit coupling and
the Coulomb term, respectively \cite{YamajiNKAI2014}. Due to the
trigonal distortion term, the spin Hamiltonian does not possess the
symmetry among spin components $S_x, S_y, S_z$; only $S_x$ and $S_y$
remain symmetric. Thus, the global symmetries of the Hamiltonian are the
$Z_2$ time-reversal symmetry, the $Z_2$ symmetry against exchange of
$S_x$ and $S_y$, and the lattice translational symmetry. By using a
vector representation of the electron creation operators at site $i$,  
${\vec{\hat{c}}_i}^\dagger =
(\hat{c}_{i,yz,\uparrow}^\dagger,\hat{c}_{i,yz,\downarrow}^\dagger,\hat{c}_{i,zx,\uparrow}^\dagger,\hat{c}_{i,zx,\downarrow}^\dagger,\hat{c}_{i,xy,\uparrow}^\dagger,\hat{c}_{i,xy,\downarrow}^\dagger)$,
the trigonal distortion term is expressed as
\begin{equation}
 \hat{H}_{\mathrm{tri}} = \sum_{i}{\vec{\hat{c}}_i}^\dagger \begin{bmatrix}
							 -\mu_{yz}&\Delta&\Delta\\
							 \Delta&-\mu_{zx}&\Delta\\
							 \Delta&\Delta&-\mu_{xy}&\\
							  \end{bmatrix}
							  \hat{\sigma}_0\vec{\hat{c}}_i,
							  \label{eq:trigonal}
\end{equation}
where $\hat{\sigma}_0$ is the $2\times 2$ identity matrix.  The {\it ab
initio} values of $\mu_{yz},\mu_{zx},\mu_{xy}$ and $\Delta$ for
$\mathrm{Na_2IrO_3}$ were estimated as $\mu_{xy}-\mu_{zx}=35$ meV, $\mu_{yz}
 \simeq \mu_{zx}$ and $\Delta=-28$ meV \cite{YamajiNKAI2014}. 

All interactions for $\mathrm{Na_2IrO_3}$ have been determined based on
the second-order perturbation theory by using the strong-coupling
expansion applied to the first principles
Hamiltonian~\cite{YamajiNKAI2014} (see Appendix
\ref{App:Hamiltonian}). Based on the exact diagonalization for small
clusters, Yamaji {\it et al} have shown that the ground state of the
Hamiltonian for $\mathrm{Na_2IrO_3}$ is expected to be a magnetically
ordered state with the zigzag configuration consistently with the
experimental observations. Within the diagonalization of small clusters,
they have also determined the phase diagram in the parameter space of
the trigonal distortion monitored around the {\it ab initio} value of
$\mathrm{Na_2IrO_3}$, and have shown that several distinct
magnetically ordered states emerge depending on the amplitude of the
trigonal distortion~\cite{YamajiNKAI2014}. However, the system size in
the previous exact diagonalization is only up to $24$ sites, and is too
small for establishing properties in the thermodynamic limit. Thus,
in order to verify the conclusion of the previous work and further
clarify the nature of the {\it ab initio} Hamiltonian that contains
anisotropic Kitaev-Heisenberg interaction, we clearly need to
investigate lager system sizes.  In the following sections, we clarify
the ground state properties in the thermodynamic limit derived from
results of much larger system sizes including the infinite-size
calculations.  We focus on the two problems; properties of the {\it ab
initio} Hamiltonian for $\mathrm{Na_2IrO_3}$ in comparison with the
experimental results and the phase diagram of the Hamiltonian obtained
by monitoring the trigonal distortion $\Delta$ away from the {\it ab
initio} value, because $\Delta$ is an experimentally tunable control
parameter by pressure or elemental substitutions.

\section{Classical approximation\label{Sec:Classical}}
We first consider the classical ground state, motivated by the fact that
the system has a magnetic order in the experimentally observed
ground state. In particular, the classical analysis provides us with
insight complementary to the quantum analysis, because it enables studies on incommensurate order, whereas it is hard to capture within
the framework of the TN or DMRG in which commensurability is assumed.

In the case of the classical Heisenberg spins, where $\vec{S}_i$ is a unit
vector with three components, the candidates of the ground state is
obtained from the Fourier transform of exchange interactions. Suppose
two spins connected by the $z$-bond on the honeycomb lattice as a
unit (see Fig.~\ref{fig:unit_cells}(c)). Because the Hamiltonian retains the translational symmetry based
on such units constituting the triangular lattice, the Hamiltonian is
diagonalized by the Fourier transform as
\begin{equation}
 \mathcal{H}_{\mathrm{cl}} = \frac{1}{2}\sum_{\bm{q}}\vec{S}_{-\bm{q}}^T
  \mathcal{J}_{\bm{q}} \vec{S}_{\bm{q}},
\end{equation}
where $\sum_{\bm{q}}$ is the summation over the wavevectors in the
Brillouin zone. $\mathcal{J}_{\bm{q}}$ is the $6\times 6$ Hermitian
matrix representing the Fourier transform of the exchange interaction
and it is given by
\begin{align}
 \mathcal{J}_{\bm{q}} &\equiv \begin{pmatrix}
			      A&B\\
			      B^\dagger&A
			     \end{pmatrix} \notag \\
 A & =			     			      2
 J^{(\mathrm{2nd})}
			      \cos
 {\bm{q}\cdot\left(\bm{a}-\bm{b}\right)} \notag \\
 B & = J_x e^{-i
			      \bm{q}\cdot\bm{a}} + J_y e^{-i
			      \bm{q}\cdot\bm{b}} + J_z
 \notag \\
			      &\qquad +J^{(\mathrm{3rd})} \left[2 \cos\bm{q}\cdot(\bm{a}-\bm{b})
			      +e^{-i\bm{q}\cdot(\bm{a}+\bm{b})} \right],
\end{align}
where $\vec{S}_{\bm{q}}$ is the Fourier transform of spins in a unit 
\begin{equation}
 \vec{S}_i \equiv
 \begin{pmatrix}
  \vec{S}_i^{(1)}\\
  \vec{S}_i^{(2)}
 \end{pmatrix}.
\end{equation}

By diagonalizing the exchange matrix $\mathcal{J}_{\bm{q}}$ numerically,
we obtain candidates of the classical ground states as the lowest eigen mode. However, note that the obtained eigenvector is not necessarily the ground state because it may not satisfy the fixed length condition
of classical spins; $|\vec{S}_i^{(1)}| = 1$ and $|\vec{S}_i^{(2)}| =
1$ for each site on the lattice. Instead, the lowest eigen mode can be regarded as the ground state in the spherical approximation $\sum_{i} |\vec{S}_i|^2 =
2N$. 
When the wavevector of the lowest eigen mode is incommensurate to
the lattice, the eigen vector generally does not satisfy the fixed length
condition. In such cases, a commensurate order, which usually satisfies the fixed
length condition, close to the incommensurate wavevector often appears as the true ground state by
recovering the fixed length condition. 

\begin{figure*}
  \includegraphics[width=16cm]{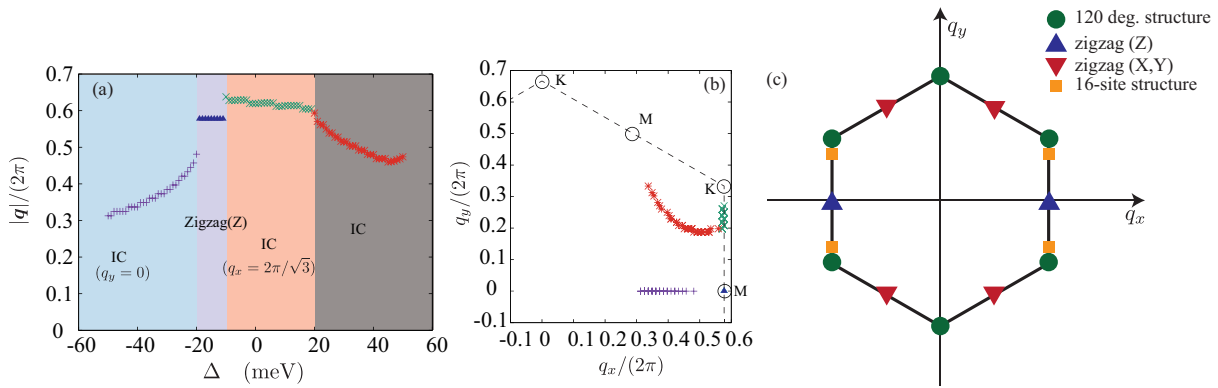} \caption{(a)
  Classical phase diagram within the spherical approximation in the
  parameter space of the trigonal distortion around the {\it ab initio}
  value. The symbols represent the amplitude of the wavevectors for the
  ground state. (b) The position of the ground-state wavevectors. The
  symbols (colors) follows the notaion in (a). The
  dashed lines represent the boundaries of the first Brillouin zone and
  the circles indicate high symmetric points. (c) The Bragg peak
  positions of several magnetic orders that appear in the present {\it ab
  initio} Hamiltonian (see also Fig.~\ref{fig:mag_fig})  \label{fig:classical_phase}}
\end{figure*}

\begin{figure}
  \includegraphics[width=8cm]{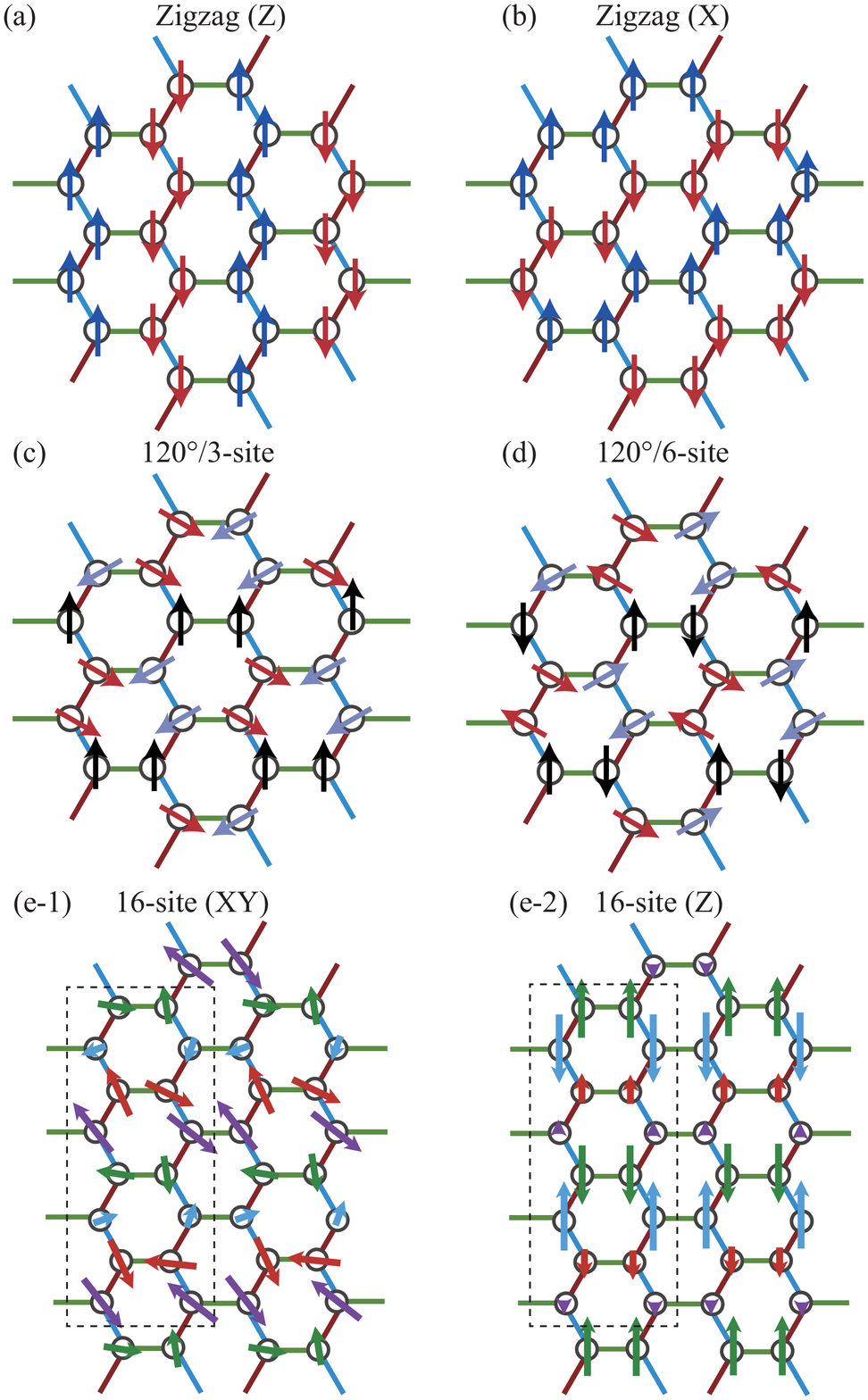} \caption{Schematic
views of magnetically ordered states emerging in the present {\it ab
initio} quantum Hamiltonian and its extensions containing Kitaev and
extended Heisenberg interactions: (a) zigzag(Z), (b) zigzag(X), (c)
$120$ degree structure with $3$-site structure, (d) $120$ degree
structure with $6$-site structure, and (e-1) $xy$ components and (e-2)
$z$ components of the $16$-site structure. For the $16$-site structure,
dashed rectangles indicate a magnetic unit cell.\label{fig:mag_fig}}
\end{figure}

In Fig.~\ref{fig:classical_phase}(a), we plotted the wavevector of the
lowest energy mode as a function of the trigonal distortion $\Delta$.
We see that in the most part of the parameter $\Delta$, 
the lowest eigen
mode appears at wavevectors incommensurate to the lattice (see
Fig.~\ref{fig:classical_phase}(b)). However, for $-20~\mathrm{meV}
\lesssim \Delta \lesssim -10~\mathrm{meV}$ we find 
that 
the commensurate
zigzag($Z$) order is the ground state, where the ferromagnetically-ordered chains consist of
$X$ and $Y$-bonds while these chains are anti-ferromagnetically coupled
by the $Z$-bonds (see Fig.~\ref{fig:mag_fig}(a)). 
Note that the {\it ab
initio} value for the distortion of $\mathrm{Na_2IrO_3}$ is $\Delta \simeq -28
\mathrm{meV}$, which is in the vicinity of the zigzag($Z$) phase
boundary. 

The obtained $\Delta$ dependence of the wave vector 
clearly indicates that the incommensurate region consists of three
distinct phases (see Fig.~\ref{fig:classical_phase} (a)). For large
negative $\Delta$, the lowest energy states are characterized by the
wavevectors $q_y = 0$ (the IC($q_y=0$) phase), which move toward the $M$
point characterizing the zigzag(Z) state. Around $\Delta = 0$, the Bragg
wave number of the lowest energy states is located on the line
connecting the $M$ and $K$ points (the IC($q_x = 2\pi/\sqrt{3}$)
phase). Note that the M point represents the zigzag phase while the K
point is the Bragg point for the 120 degree phase.  Finally, for large
$\Delta$ the lowest energy state appears inside the first Brillouin zone
(the IC phase). Note that at the phase boundary between the zigzag(Z)
and the IC ($q_x = 2\pi/\sqrt{3}$) phases, the characteristic wavevector
discontinuously changes from the $M$ point to a vicinity of the $K$
point. Thus, at the boundary the phase transition is expected to be of
the first order. On the other hand, the wavevector changes continuously
without a jump between IC($q_y = 0$) and the zigzag($Z$), and also
between the IC($q_x = 2\pi/\sqrt{3}$) phase and the IC phase. Although
the wavevector looks varying steeply in the former case, it is likely to
be continuous because the wavevector looks naturally connected to the $M$
point. For these boundaries, the phase transitions could be continuous.

Although the classical ground state within the spherical
approximation is characterized by the incommensurate wavenumber (IC
($q_x = 2\pi/\sqrt{3}$) phase), it is close to the commensurate $120$
degree structure near the phase boundary to the zigzag phase.  In fact,
it was claimed to be the ground state in a simple classical model
proposed by Rau {\it et al.,} \cite{RauLK2014,RauK2014arXiv}: They
considered the classical Kitaev-Heisenberg model with an additional
$I_1$ term, {\it e.g.}  $S_xS_y$ interaction on $z$-bonds. In their
model, the $120$ degree structure appears as the ground state for a
region with the antiferromagnetic Kitaev coupling and the
antiferromagnetic Heisenberg coupling with finite $I_1$ term. Although
they have not discussed details of this $120$ degree structure, in their
model, it is actually degenerate with states that have any relative
angles between the clusters consisting of three neighboring sites of a
site, where the three sites are aligned mutually with $120$ degrees (see
examples in Figs.~\ref{fig:mag_fig}(c) and (d)).  When we set zero
relative angle, we obtain a state that has a unit-cell containing $3$
sites, which we call $3$-site structure (Fig.~\ref{fig:mag_fig}(c)). (A
similar terminology will be used below for larger unit-cell
structures.). On the other hand, the relative angle of $180$ degrees
makes a $6$-site structure (Fig.~\ref{fig:mag_fig}(d)). Furthermore any
value of the relative angle produces a ground state.  Note that the
classical ground state within the spherical approximation does not
contain the $120$-degree commensurate phase, but the IC ($q_x =
2\pi/\sqrt{3}$) order replaces it, although they are close in energy.
 Such degeneracy among various
$120$-degree orders  is often
lifted by thermal or quantum fluctuations by the so called
order-by-disorder mechanism \cite{VillainBCC1980}. Actually, our
classical Monte Carlo calculation of the Rau's model shows that either
the $3$-site or the $6$-site $120$- degree commensurate order is
selected at finite temperatures depending on the sign of $K$ and $I_1$.  Thus, we expect that if the commensurate $120$- degree
order is stabilized in stead of the IC ($q_x = 2\pi/\sqrt{3}$) order in
the present {\it ab initio} Kitaev-Heisenberg Hamiltonian with $S=1/2$
quantum spins, it could be either the $3$-site or the $6$-site
structures, as reported in previous $24$-site exact-diagonalization
calculations\cite{YamajiNKAI2014}.

In addition, we also expect that other types of magnetic order
may be stabilized because incommensurate wavevectors plotted in
Fig.~\ref{fig:classical_phase} are close to some commensurate values. For
example, in the IC($q_x = 2\pi/\sqrt{3}$) region, the wavevectors are
located in the vicinity of $\vec{q}/2\pi =(1/\sqrt{3},1/4)$ and
$\vec{q}/2\pi =(1/\sqrt{3},1/5)$ which corresponds to a $16$-site
structure and a $20$-site structure, respectively. Unfortunately, these
large unit-cell structure are not fitted to $24$-site cluster used in
previous ED calculation\cite{YamajiNKAI2014}. In order to investigate stability of such
structures, we need larger unit-cells beyond ED.

\section{Ground state properties\label{Sec:GroundState}}
\subsection{Methods}
In order to investigate the ground state property of the Hamiltonian
in the presence of strong quantum fluctuations, we conduct three types of
numerical calculations based on the ED, 
two-dimensional DMRG and tensor network based methods.

Our ED calculations were done up to $32$-site based on the Lanczos
algorithm. A part of the results has already been reported in
Ref.~\onlinecite{YamajiNKAI2014}. In order to investigate larger
systems, we use a two-dimensional DMRG method where we represent the
ground state wave function as a matrix product states (MPS) and
variationally optimize the wave-function parameters so as to minimize
the energy \cite{ShinjoST2015}.  In the DMRG calculation, we investigate
$L_x \times L_y$ lattice systems with periodic boundary conditions along
both of $L_x$ and $L_y$ directions (see
Fig.~\ref{fig:unit_cells}(a)). We keep $1000$ states in DMRG processes
and perform more than 10 sweeps, resulting in a typical truncation error
$10^{-5}$ or smaller.

The ED and DMRG are quite accurate for finite systems. The maximum size they
can treat is, however, restricted to about $100$ sites, which are a little
too small to clarify the thermodynamic properties beyond reasonable
doubt. In order to investigate the property in the thermodynamic limit
further, we also conduct recently developed tensor network methods
which can treat infinite-size system directly. Here we use
the tensor network ansatz for infinite-size systems so called infinite Projected Entangled Pair State (iPEPS)
\cite{VerstraeteC2004,VerstraeteC2004arXiv,JordanOVVC2008} or infinite
Tensor Product State (iTPS) \cite{MartinRS2001,NishinoHOMAG2001} as the
ansatz of the ground state wave function.  We assume infinitely repeated
$L_x \times L_y$ unit-cell structure, {\it i.e.} $L_x \times L_y$
independent tensors with bond-dimensions $D$, which is the same shape
with the lattice shape used in DMRG (Fig.~\ref{fig:unit_cells}). Note
that the unit-cell structure used in iPEPS allows the  ground state that spontaneously breaks the lattice
translational symmetry  in the thermodynamic limit with the  periodicity taken into account up to the unit cell size.
Thus, even if we
assume a finite $L_x \times L_y$ unit-cell structure, a wave function
represented by iPEPS is that of the infinite system without any finite
size boundary effects. For the optimization of the tensors, we use the
imaginary-time evolution with so called the simple update technique
\cite{JiangWX2008}, which is extended to treat the second- and third-neighbor
interactions (see Appendix \ref{App:PEPS}). Reliability of the simple
update optimization is demonstrated by comparing with the full update
\cite{JordanOVVC2008,OrusV2009} for the nearest-neighbor {\it ab
initio} Hamiltonian in Appendix \ref{App:PEPS_NN}. After obtaining optimized
tensors, we calculated physical quantities by using the corner transfer
matrix method \cite{Baxter1968,Baxter1978,Baxter_book,NishinoO1998,OrusV2009,CorbozWV2011,CorbozRT2014}. In the
following calculations, we use the bond dimension $D\le 9$ for the case
of the {\it ab initio}
value of $\mathrm{Na_2IrO_3}$ and $D \le 6$ for the case of
the trigonal distortion controlled away from the {\it ab initio} value.

\subsection{Ground state of $\mathrm{Na_2IrO_3}$}
First, we examine the ground state of the Hamiltonian at the {\it ab
initio} matrix elements in Eq.~\eqref{eq:Hamiltonian} for
$\mathrm{Na_2IrO_3}$ obtained by a first principles calculation
\cite{YamajiNKAI2014}. Based on the second order perturbation theory, we
can evaluate the exchange interactions as a function of $\Delta $
\cite{YamajiNKAI2014}. For $\mathrm{Na_2IrO_3}$, the {\it ab initio}
value of the trigonal distortion was calculated as $\Delta = -28 $ meV,
and estimated interactions of the Hamiltonian are listed in Table
\ref{table:exchange} (see also Appendix \ref{App:Hamiltonian}).  At this value, $24$-site ED calculation predicts
that the ground state is the zigzag($Z$) state where the
ferromagnetically-ordered chains consist of $X$ and $Y$-bonds while
these chains are antiferromagnetically coupled by the $Z$-bonds (see
Fig.~\ref{fig:mag_fig}(a))\cite{YamajiNKAI2014}.
\begin{table}
\begin{tabular}{c|c c c c c c}
\hline
\hline
 $J_{X,Y}~(\mathrm{meV})$&$K'$ & $J'$&$J''$&$I_1'$&$I_2'$&$I_2''$ \\
\hline
 & $-23.9$&$2.0$&$3.2$&$1.8$&$-8.4$&$-3.1$\\ 
\hline
\hline
 $J_Z~(\mathrm{meV})$& $K$&$J$&$I_1$&$I_2$\\
\hline
&$-30.7$&$4.4$&$-0.4$&$1.1$\\
\hline
\hline
 $J_2~(\mathrm{meV})$& $K^{(2nd)}$&$J^{(2nd)}$&$I_1^{(2nd)}$&$I_2^{(2nd)}$\\
\hline
&$-1.2$&$-0.8$&$1.0$&$-1.4$\\
\hline
\hline
 $J_3~(\mathrm{meV})$& $J^{(3rd)}$\\
\hline
 &$1.7$\\
\hline
\hline
\end{tabular}
\caption{Kitaev and extended Heisenberg exchange interactions derived from the second-order
 perturbation theory applied to the {\it ab initio} 
 Hamiltonian at $\Delta = -28$ meV~\cite{YamajiNKAI2014}} \label{table:exchange}
\end{table}

\begin{figure}
  \includegraphics[width=8cm]{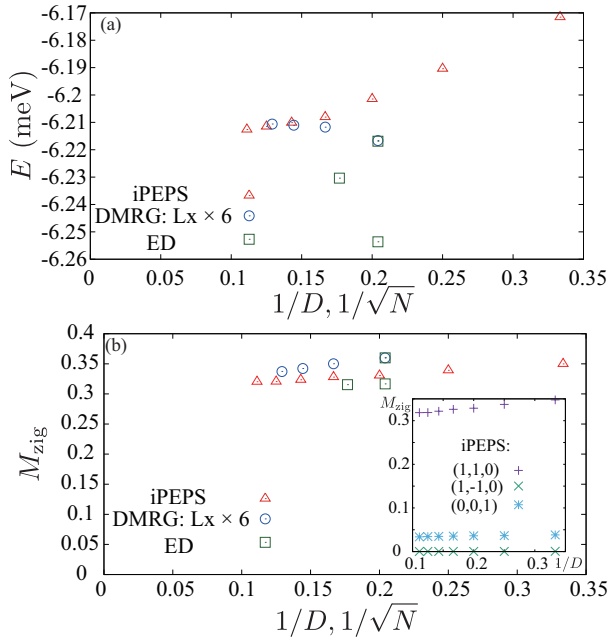}
  \caption{Bond-dimension ($D$) or system size ($N$) dependences of the
  ground-state energy (a), and zigzag(Z) order parameter (b) calculated
  by ED, DMRG and iPEPS for the {\it ab initio} Hamiltonian of
  $\mathrm{Na_2IrO_3}$, where the Kitaev and extended Heisenberg
  interactions are derived from the {\it ab initio} value $\Delta = -28
  {\mathrm{meV}}$. The inset of (b) shows the amplitude of $\langle
  \vec{M}_{{\rm zig}(Z)}\rangle $ projected onto $(x,y,z) = (1,1,0),
  (1,-1,0)$ and $(0,0,1)$ directions obtained by iPEPS ($D=9$).  \label{fig:abinitio}}
\end{figure}

In Fig.~\ref{fig:abinitio}(a), we show the energies calculated by ED,
DMRG, and iPEPS as a function of the $1/\sqrt{N}$ (ED, DMRG) or $1/D$
(iPEPS). In the case of iPEPS, we used $L_x \times L_y = 4\times 6$ unit
cell. We see that the energies calculated by different methods are
consistent with each other and they seem to reach a common value, $E
\simeq -6.22$ meV in the limit of $D, N \to \infty $. 

In order to further clarify the nature of the ground state, we plot in
Fig~\ref{fig:abinitio}(b) the order parameter of the zigzag(Z) state
defined as
\begin{equation}
 \vec{M}_{\mathrm{zig(Z)}} \equiv \frac{1}{2}(\vec{\sigma}_1 -\vec{\sigma}_2),
\label{eq:order_zigzag_z}
\end{equation}
where $\sigma_{\alpha}$ ($\alpha = 1,2$) represents the average of spins
over the equivalent sites in zigzag(Z) state (see
Fig.~\ref{fig:mag_fig}(a)).  In the cases of ED and DMRG we plot
$M_{\mathrm{zig(Z)}} =
\sqrt{\langle\vec{M}_{\mathrm{zig(Z)}}^2\rangle}$, while in the case of
iPEPS we plotted $M_{\mathrm{zig(Z)}} =
\sqrt{\langle\vec{M}_{\mathrm{zig(Z)}}\rangle^2}$. Note that these two
definitions should reach the same thermodynamic limit.  One can clearly
see that $M_{\mathrm{zig(Z)}}$ takes a large finite value and remains
nonzero in the limit of $D, N \to \infty$, indicating that zigzag(Z)
state is stabilized as the ground state in the thermodynamic
limit. Thus, the previous proposition that the ground state of the {\it
ab initio} Hamiltonian of $\mathrm{Na_2IrO_3}$ is the zigzag($Z$) state
has been established beyond reasonable doubt. In the inset of
Fig.~\ref{fig:abinitio}(b), we also plot the amplitude of $\langle
\vec{M}_{{\rm zig}(Z)}\rangle $ projected onto $(x,y,z) = (1,1,0),
(1,-1,0)$ and $(0,0,1)$ directions obtained by iPEPS. We see that the
$(1,1,0)$ component is dominant rather than $(1,-1,0)$ and $(0,0,1)$
components. Similar results have been reported based on the pinning
field analysis of ED in the previous study \cite{YamajiNKAI2014}. Thus
the ordered moment in this zigzag(Z) state is nearly along $(1,1,0)$
direction in the effective spin basis, although it has also weak ($<
15\%$) $(0,0,1)$ component.

From x-ray resonant-magnetic-scattering experiments, the ordered moment
in the low temperature phase of $\mathrm{Na_2IrO_3}$ was estimated to be
located in the $ac$ plane \cite{Liu2011,LoveseyD2012}. Because 
$(1,1,0)$ direction in the effective spin model is converted to
$(1,1,0)$ direction in the real space\cite{YamajiNKAI2014}, which is on
the $ac$ plane of $\mathrm{Na_2IrO_3}$, the predicted ordered spin
direction is consistent with the experimental observation. 
On the other hand, further quantitative analysis of the experimental
data suggested that the ordered moment was almost parallel to the $a$
axis\cite{Liu2011,LoveseyD2012}, which was $(1,1,-2)$ directions in the
real space. Thus, ordered spin direction $(1,1,0)$ predicted from the
{\it ab initio} Hamiltonian does not completely match the experimental
observation. In order to reproduce the ordered-moment direction more
precisely, we might need to take into account the coupling between the
honeycomb layers, which is ignored in the present {\it ab initio}
Hamiltonian\cite{YamajiNKAI2014}.

\subsection{Phase diagram for parameters away from the {\it ab initio} trigonal distortion}
Next, we discuss the ground-state phase diagram in the parameter space
of the trigonal distortion monitored around the {\it ab initio} value
$\Delta = -28$ meV. Once we set $\Delta$, one can calculate the exchange
couplings as a function of $\Delta $ through the second-order
perturbation theory \cite{YamajiNKAI2014}. As expected from the
classical analysis, we need to keep in mind that magnetically ordered
ground states stabilized by controlling the trigonal distortion are
subject to have large unit cells including periodicity that is
incommensurate to the lattice. In order to obtain the true ground state,
we examine the dependence on the system size (DMRG) and on the assumed
unit cell size and structure (iPEPS). We then take the lowest energy state
among various choices as the ground state.

\begin{figure}
  \includegraphics[width=8cm]{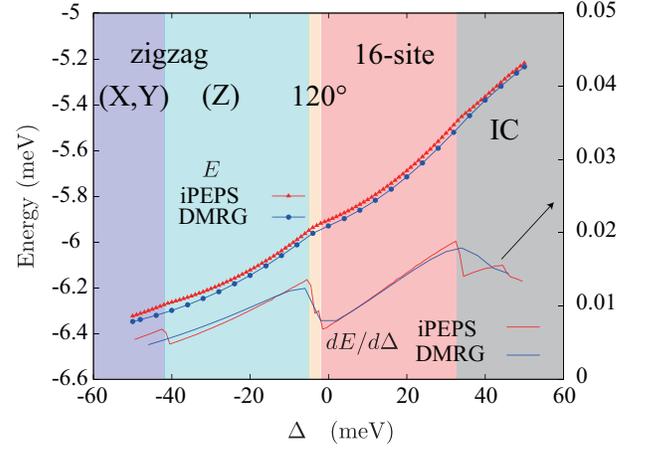} \caption{Phase
  diagram in parameter space of trigonal distortion around {\it ab
  initio} Hamiltonian represented by Kitaev and extended Heisenberg
  interactions for $\mathrm{Na_2IrO_3}$. Red triangles and blues circles
  represent the energies calculated by iPEPS with $D=6$ and DMRG for the
  $6\times8$ lattice, respectively. In the case of iPEPS we calculated
  the energy for $4\times4$, $6\times8$, $6\times 10$, and $8\times12$
  unit cells and took the lowest energy state among them as the ground
  state. The derivatives of the energies with respect to $\Delta$ are
  also shown by solid curves without symbols to gain insight into the
  positions of the phase transitions.\label{fig:phase_diagram}} 
\end{figure}

In Fig.~\ref{fig:phase_diagram}, we show the phase diagram  that is determined
from consistent results of DMRG and iPEPS calculations, together with the energies and
their derivatives calculated by iPEPS and DMRG. In the phase diagram, we
find four types of magnetically ordered states.

For $ \Delta \lesssim -3~\mathrm{meV}$, the zigzag state is
stabilized. The zigzag phase is separated in two types depending on the
direction of ferromagnetically-ordered chains. For $\Delta \lesssim
-44~\mathrm{meV}$ the ferromagnetically-ordered chains are perpendicular
to the $X$-bond or to the $Y$-bond (zigzag($X$) and zigzag($Y$) states,
respectively), while for $ -44~\mathrm{meV}\lesssim \Delta \lesssim
-3~\mathrm{meV}$ they are perpendicular to the $Z$-bond (See
Figs.~\ref{fig:mag_fig} (a),(b)). This zigzag($Z$) phase contains the
{\it ab initio} value of $\mathrm{Na_2IrO_3}$, $\Delta = -28$meV. When
we increase $\Delta$, the $120$ degree structure is stabilized for $
-3~\mathrm{meV}\lesssim \Delta \lesssim -1~\mathrm{meV}$.  Whereas the
previous $24$-site ED calculation suggested that the $120$ degree
structure survives for larger trigonal distortion $\Delta < 40 \mathrm{meV}$
\cite{YamajiNKAI2014}, the new results based on iPEPS and DMRG reliably
show that the $16$-site structure characterized by the wavevector
$\bm{q}/2\pi=(1/\sqrt{3}, 1/4)$ (see Fig.~\ref{fig:classical_phase}(c))
is stabilized for $-1~\mathrm{meV} \lesssim \Delta \lesssim
30~\mathrm{meV}$.
Finally, for even larger $\Delta$ values ($\Delta\gtrsim
30~\mathrm{meV}$), large unit-cell magnetic structures appear as the
ground state.  As is indicated by the anomaly around $45~\mathrm{meV}$
in the iPEPS result of $dE/d\Delta$, there are at least two types of
states in this region. These two states have the $48$-site magnetic
unit-cells with different shapes, which are equal to unit-cell size used
in the iPEPS calculation ($6\times 8$) or a half of the size ($8\times
12$). Comparing the two states with the classical states in Fig.~\ref{fig:classical_phase} (a),
we speculate that the ground state in this region is incommensurate to
the periodicity of the honeycomb lattice in the thermodynamic limit. The
reason why only the two states are observed may be attributed to the
limited sizes allowed for the iPEPS unit cell.

\begin{figure}
  \includegraphics[width=8cm]{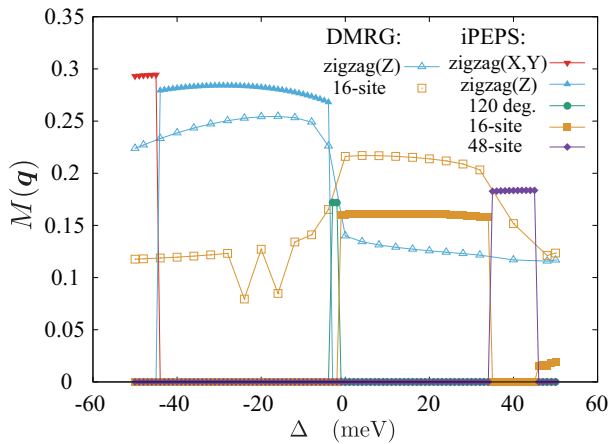} \caption{Order parameters
  $M(\bm{q})$ as a function of trigonal distortion. For the nature of
  the wavevector characterizing each state, see the main text (See also
  Fig.~\ref{fig:classical_phase}(c) for the Bragg wave number of each
  order). In the case of DMRG, we use $6\times8$ lattice. Thus, we only
  plot the order parameter commensurate to $6\times8$ lattice in the
  case of DMRG. \label{fig:SQ}} 
\end{figure}

In order to see these magnetic orders clearly, we define the order
parameter through the spin structure factor as
\begin{equation}
 M(\bm{q}) \equiv \sqrt{\frac{1}{N}\sum_{\gamma =x,y,z} \sum_{i=1}^N 
\langle \hat{S}_0^\gamma\hat{S}_i^\gamma\rangle
\cos(\bm{q}\cdot\bm{r}_i)}.
\end{equation}
In the case of iPEPS, we use  
\begin{equation}
 M(\bm{q}) \equiv \sqrt{\sum_{\gamma =x,y,z}\left| \frac{1}{N}\sum_{i=1}^{N} 
\langle \hat{S}_i^\gamma\rangle e^{i\bm{q}\cdot\bm{r}_i} \right|^2}.
\end{equation}
Because the wave function obtained by iPEPS is that of the infinite
system, we calculate $M(\bm{q})$ of iPEPS approximately by using the $96
\times 96$ finite lattice. In Fig.~\ref{fig:SQ}, we plot
$M(\bm{q})$s corresponding to zigzag($Z$), zigzag($X$) (and
zigzag($Y$)), $120$ degree structure and $16$-site order together with
the $48$-site order representing the incommensurate region in the phase
diagram. The Bragg wavevectors for each state is given as
\begin{equation}
 \frac{\bm{q}}{2\pi} = 
\begin{cases}
 \left(\frac{1}{2\sqrt{3}},\frac{1}{2}\right), \left(-\frac{1}{2\sqrt{3}},\frac{1}{2}\right) & \text{(zigzag($X,Y$))}\\
 \left(\frac{1}{\sqrt{3}},0\right) & \text{(zigzag($Z$))}\\
 \left(\frac{1}{\sqrt{3}},\frac{1}{3}\right) & \text{($120$ deg.)}\\
 \left(\frac{1}{\sqrt{3}},\frac{1}{4}\right) & \text{($16$-site)}\\
 \left(\frac{3}{4\sqrt{3}},\frac{1}{3}\right) & \text{($48$-site)}
\end{cases}.
\end{equation}
One can see that corresponding order parameters become finite in each
phase, which ensures the stability of the identified phases. Note that
the order parameters at the phase boundaries remain at large nonzero
values before the transition to zero indicating the first order nature
of the phase transitions. In case of the DMRG, the finite-size effects
smears the jump to some extent.

In the following, we investigate details of each phase in the
phase diagram. 

\subsubsection{zigzag phase}
\begin{figure*}
  \includegraphics[width=16cm]{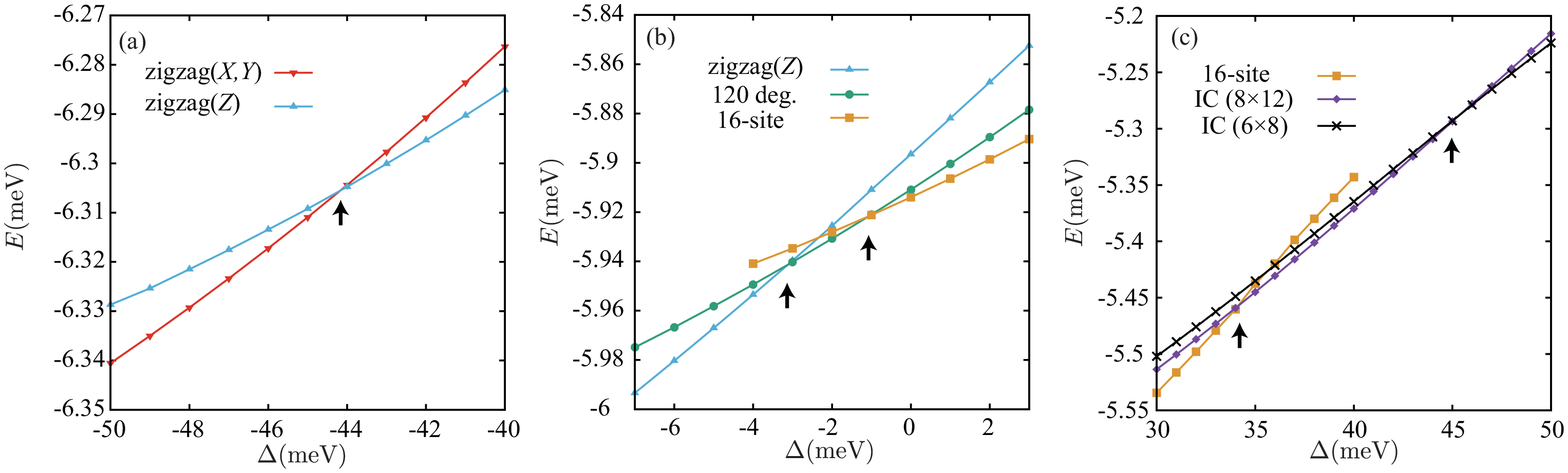} \caption{(Color
  online) Energies obtained by iPEPS in the vicinity of phase
  boundaries. (a) Phase boundary between the zigzag($X$, $Y$) and
  zigzag($Z$) phases. (b) Phase boundaries of the $120$ degree phase. (c)
  Phase boundaries of the IC phase. The arrows indicate the positions of
  the phase boundaries.\label{fig:energy_cross}}  
\end{figure*}
As we have shown in the phase diagram, two types of zigzag states are
stabilized in the negatively large $\Delta$ region. In the vicinity of
the phase boundary, we can obtain both of the zigzag(X,Y) and the
zigzag(Z) states depending on the unit-cell structures 
and/or the initial conditions of the tensors in the iPEPS, at least as a metastable states. Thus, we can locate the first-order phase boundary at the point
where the two energy curves cross each other. In
Fig.~\ref{fig:energy_cross} (a), we plot the energies of the zigzag(X,Y)
state and zigzag(Z) state obtained by the iPEPS. We see a clear energy
crossing around $\Delta = -44$ meV indicating the first-order phase
transition between the zigzag(X,Y) phase and the zigzag(Z) phase.

In the DMRG for $6\times 8$ cluster, there is no clear anomaly
around $\Delta = -44$ meV in the energy and the order parameters (see
Figs.~\ref{fig:phase_diagram} and \ref{fig:SQ}). Because $6\time 8$
cluster does not fit the zigzag(X,Y) structure, the zigzag(Z) state
is probably stabilized in wider region than the iPEPS.

Note that the degeneracy of the ground states is different between the
zigzag(X,Y) and the zigzag(Z) phases;  four-fold degeneracy for the zigzag(X,Y) state, while two-fold for the zigzag(Z) state. Thus we expect
finite-temperature phase transitions with the $Z_4$ symmetry breaking for the zigzag($X,Y$) phase,  distinct from the $Z_2$
symmetry breaking for the zigzag($Z$) state.

\subsubsection{$120$ degree phase}
Next we focus on the phase with the $120$ degree structure. In
Fig.~\ref{fig:energy_cross}(b), we plot the energies of the zigzag(Z),
the $120$ degree, and the $16$-site states as a function of $\Delta$. For $-3~\mathrm{meV} \lesssim \Delta \lesssim -1 \mathrm{meV}$,
the $120$ degree state has the lowest energy. Although, the range of
the $120$ degree structure is largely reduced from the previous estimate
based on the $24$-site ED calculation \cite{YamajiNKAI2014}, the data
show that the $120$ degree structure phase survives.

Although the $120$ degree structure does not fit $6\times 8$ cluster
used in DMRG, we observed that in DMRG for $8\times 6$ cluster the
$120$ degree structure appears and around $\Delta = 0$ meV its energy is
lower than that of $6\times 8$ cluster. This observation also indicates
that around $\Delta = 0$ meV, the $120$ degree structure is stabilized
rather than the $16$-site structure. 

As we mentioned in Sec.~\ref{Sec:Classical}, in the case of the simple
Rau's model, the classical $120$ degree state is highly degenerated
including $3$-site and $6$-site states in Fig.~\ref{fig:mag_fig}
(c,d). In the case of the present {\it ab initio} Hamiltonian, the $120$
degree state is expected to be slightly distorted from $120$ degrees
structure by varying the canting angle because of the anisotropy. Then
the quantum fluctuation and the distortion would lift the degeneracy
between $3$-site and $6$-site states, and one of them could be realized
as the ground state.  Although
it is difficult to determine which structure is actually realized in the
ground state from finite-size calculations that prohibit spontaneous
symmetry breaking, we can investigate the structures of infinite system
using iPEPS where we observe the spontaneous symmetry breaking measured
by a nonezero local magnetization.

In order to investigate the magnetic structure, we study the relative
angle $\phi$ between two spins connected by $z$-bond, which is defined
as
\begin{equation}
 \cos (\phi) \equiv \frac{\langle \vec{S}^{(1)} \rangle \cdot \langle
  \vec{S}^{(2)} \rangle \cdot }{|\langle \vec{S}^{(1)} \rangle||\langle
  \vec{S}^{(2)} \rangle|}.
\end{equation}
In the ideal $120$ degree structure, $\cos(\phi)$ takes three values depending on the
position.
For the $3$-site and the $6$-site structures,
they are
\begin{equation}
 \cos(\phi) = \begin{cases}
	       (1, -1/2, -1/2) & 3\text{-site}\\
	       (-1, 1/2, 1/2) & 6\text{-site}.
	      \end{cases}
\end{equation}
In the iPEPS, we obtained two sets of $\cos(\phi)$ depending on the
initial conditions: They are $\cos\phi = (-0.98, 0.75, 0.58)$ and
$\cos\phi = (1.00, -0.25, -0.26)$ at $\Delta = -2$ meV. Based on the
comparison with the expected values of the $3$-site and the $6$-site
structures, we interpret that the states obtained by iPEPS correspond to
$3$-site and $6$-site states with distortion caused by the
anisotropy. The energy of the expected $6$-site state is slightly
lower than that of the $3$-site state. However, the energy difference
is only $\Delta \simeq 0.0003$ meV which seems to be smaller than the
numerical errors that arise from the imaginary-time evolution with
finite time steps \cite{JiangWX2008} and the approximate contraction of
the infinite tensor
network\cite{OrusV2009,CorbozWV2011,CorbozRT2014}. Thus, based on the
present numerical data, we only conclude that the $3$-site state or
the $6$-site state are realized as the ground states in the $120$ degree
phase.

Both of the $3$-site and $6$-site $120$ degree states are six-fold
degenerate: $Z_3$ from the lattice translation and $Z_2$ from the time
reversal symmetries. Thus, one can expect a finite-temperature phase
transition with breaking of the $Z_6$ symmetry, which is usually a
successive Berezinskii-Kosterlitz-Thouless (BKT) transitions. By decreasing the
temperature from the high temperature paramagnetic phase, a BKT transition
occurs at $T=T_{c1}$ and quasi long-range ordered phase of an emergent
$U(1)$ symmetry appears in $T_{c2} < T <T_{c1}$. Below the second BKT
transition temperature $T_{c2}$, the magnetic long-range order with
breaking $Z_6$ symmetry is expected to be stabilized \cite{JoseKKN1977}.
\subsubsection{$16$-site phase}
Based on the DMRG and iPEPS, we found that the $16$-site state, which
has not been reported in the previous analyses, is stabilized in a wide
region of the phase diagram. Here we investigate the magnetic structure
of the $16$-site state.

As we mentioned, the $16$-site state is characterized by the wavevector
$\bm{q}^*/2\pi=(1/\sqrt{3}, 1/4)$. Using the wavevector $\bm{q}^*$, the
local magnetizations obtained from iPEPS are well reproduced by
\begin{equation}
 \langle \vec{S}^{(1)}(\bm{r})\rangle = \begin{pmatrix}
					 r_{xy} \cos(\bm{q}\cdot \bm{r} + \theta + \alpha)\\
					 r_{xy} \cos(\bm{q}\cdot \bm{r} - \theta + \alpha)\\
					 r_z \cos(\bm{q}\cdot \bm{r} + \alpha)
			 \end{pmatrix}
\end{equation}
and 
\begin{equation}
 \langle \vec{S}^{(2)}(\bm{r}) \rangle = \begin{pmatrix}
					  r_{xy} \cos(\bm{q}\cdot \bm{r} - \theta + \alpha)\\
					  r_{xy} \cos(\bm{q}\cdot \bm{r} + \theta + \alpha)\\
					  r_z \cos(\bm{q}\cdot \bm{r} + \alpha)
			 \end{pmatrix},
\end{equation}
where $\vec{S}^{(1)}(\bm{r})$ and $\vec{S}^{(2)}(\bm{r})$ are two spins
in the unit defined in Fig.~\ref{fig:unit_cells}(c) located at
$\bm{r}$. Note that the amplitudes and the phases $r_{xy}$, $r_z$,
$\theta$, and $\alpha$ depend on $\Delta$. For a better
understanding of the complex spin structure, we plot a schematic view of this
$16$-site structure in Figs.~\ref{fig:mag_fig}(e) and (f). We found that
this spin structure is consistent with the eigenvectors obtained from the
classical analysis described in Sec.\ref{Sec:Classical}.

Although the classical analysis totally ignores the quantum fluctuation effects, it still offers  
insight into the reason why the $16$-site state has lower energy than the
$120$ degree state. In the region of $-1 \mathrm{meV} \lesssim \Delta
\lesssim 20 \mathrm{meV}$, the lowest-energy mode (wavevector) in the classical
analysis is closer to the $q_y/(2\pi) = 1/4$ (the $16$-site) than the
$q_y/(2\pi) = 2/3$ (the $120$ degree) (see
Fig.~\ref{fig:classical_phase}(a,b)). Thus, the $16$-site state is more
favorable than the $120$ degree state. This simple interpretation also
explains why the $120$ degree structure is stabilized for $\Delta
\lesssim -1 \mathrm{meV}$: The lowest-energy mode of the classical analysis shows that the wavenumber approaches  $q_y/(2\pi) = 2/3$ in this region.

Because the lowest modes are also close to $\bm{q}/(2\pi) =
(1/\sqrt{3}, 1/5)$, one might speculate that a state characterized by
this wavevector could be realized. However, that is not the
case. Although we also calculated the energy using the iPEPS with $6 \times
10$ unit-cell, which is compatible with $\bm{q}/(2\pi) = (1/\sqrt{3}, 1/5)$, the
energy was higher than that of the $16$-site state. In this
model, the state with $\bm{q}/(2\pi) = (1/\sqrt{3}, 1/5)$ does not appear as the
ground state.

The $16$-site state has eight-fold degeneracy: $Z_4$ from the lattice
translation and $Z_2$ from the time reversal symmetries. In this case,
we again expect a successive BKT transitions at nonzero temperatures
with breaking of the $Z_8$ symmetry similar to the case of $Z_6$
symmetry \cite{JoseKKN1977}.

\subsubsection{Incommensurate phase}
Finally, we investigate the incommensurate phase. In
Fig.~\ref{fig:energy_cross} (c), we show the energies obtained by iPEPS
around the incommensurate phase. In the incommensurate phase, we
obtained two types of large unit-cell structure depending on the
unit-cell shape used in the iPEPS. In addition to the phase transition
between the $16$-site phase and these large unit-cell states around
$\Delta \simeq 35$ meV, one can see the energy crossing between two
distinct large unit-cell states around $\Delta \simeq 45$ meV indicated
by an arrow. Although the second energy crossing may represent a phase
transition, it could instead be ``finite size effects'' due to finite
unit-cells used in the iPEPS. Actually, in the classical analysis, the
characteristic wavevectors of the lowest-energy mode continuously shift
for $\Delta \gtrsim 35$ meV. It suggests the existence of the
incommensurate phase in this region, the possibility of which in the
quantum case is not excluded from the present analysis. Thus, we
tentatively speculate that these two large unit-cell structures are a
part of the incommensurate phase, where the Bragg wavenumber
continuously moves with $\Delta$ in the thermodynamic limit.

\section{Conclusions\label{Sec:Conclusions}}
In this paper, we have investigated the ground state properties of the
realistic effective Hamiltonian for $\mathrm{Na_2IrO_3}$. Based on the
three numerical methods, ED, DMRG and iPEPS, we have firmly established
that the ground state of the {\it ab initio} Hamiltonian for $\mathrm{Na_2IrO_3}$ is the zigzag($Z$) state, in agreement with the
experiment. In zigzag($Z$) state, ferromagnetically-coupled chains are
perpendicular to the $z$-bond and ordered spin moments are on the $ac$
plane of $\mathrm{Na_2IrO_3}$. These features are also consistent
with the experimental observation \cite{Choi2012,Ye2012,Liu2011,LoveseyD2012}.

On the other hand, the direction of the ordered moment predicted from
our {\it ab initio} Hamiltonian, which is nearly parallel to t
$(x,y,z) = (1,1,0)$ direction, does not match the analysis of
experimental data implying the moment nearly parallel to the $a$ axis
(the $(1,1,-2)$ direction) \cite{Liu2011,LoveseyD2012}. In order to
reproduce the precise direction, the weaker interactions ignored in the
present {\it ab initio} Hamiltonian, such as the couplings between
honeycomb layers, could be important. However, the experimental data are
not from the single crystal and more definite determination of the
direction is also desired.

We have also determined the ground state phase diagram of the
Hamiltonian when the trigonal distortion $\Delta$ is monitored as a
control parameter away from the {\it ab initio} value for
$\mathrm{Na_2IrO_3}$. We have found, at least, five distinct
magnetically ordered phases:the zigzag($X,Y$), the zigzag($Z$),the $120$
degree structure, the $16$-site structure, and the presumably
incommensurate phase that appears as $48$-site states in the present
calculation. For large negative $\Delta$ region, zigzag($X,Y$) states
are stabilized. When we increase $\Delta$, the direction of the
ferromagnetically-coupled chain in the zigzag state rotates $120$ degree
forming zigzag($Z$) state around $\Delta \simeq -44$ meV. In the middle
of the phase diagram $120$ degree structure appears in the narrow
region. We have identified it as the $3$-site and/or the $6$-site order
according to the result of the iPEPS calculations for infinite
systems. When we increase $\Delta$ toward positive values, the $16$-site
phase is stabilized. Although the previous $24$-site ED calculation has
not identified this phase, we have found that it has a lower energy than
that of the $120$ degree structures in a wide region based on the DMRG
and iPEPS calculations. For large positive $\Delta$, we have found a
presumable incommensurate state. Although it is difficult to prove the
``true'' incommensurate nature because of the limitation of finite-size
systems (ED, DMRG) or the number of the independent tensors (iPEPS), the
results indicate that its magnetic unit cell is at least larger than
that consisting of $48$ spins.  Since the trigonal distortion may be
selectively controlled by substitution of elements in
$\mathrm{Na_2IrO_3}$, the present results give a useful guideline to
understand possible rich phase diagram based on the combined effort of
the reliable {\it ab initio} approach and experimental progress.

The accuracy and reliability of the {\it ab initio} Hamiltonian has been established in the present study through the detailed comparison with the experimental indications: The present study by combining three independent and accurate numerical algorithms has enabled a reliable approach to the thermodynamic limit by keeping high accuracy of the result. 

An interesting question untouched in the present study on the {\it ab initio} 
Hamiltonian and left for future studies is how we can approach the spin-liquid state starting from
the magnetic $\mathrm{Na_2IrO_3}$ under the normal pressure. Based on
$24$-site ED calculation, Yamaji {\it et al} pointed out that a
lattice expansion from $\mathrm{Na_2IrO_3}$ may stabilize a Kitaev-type spin-liquid
state \cite{YamajiNKAI2014}. Since it was already shown that the iPEPS is able to describe the Kitaev spin liquid state reliably~\cite{IreguiCT2014}, searching and designing the spin liquid state by using the combined DMRG or PEPS on the realistic and first-principles framework
is an intriguing future subject.  

\begin{acknowledgements}
 We thank S.~Todo, T.~Suzuki, and K.~Harada for fruitful
 discussions. The computation in the present work is partly performed on
 computers at the Supercomputer Center, ISSP, University of Tokyo. The
 numerical package $\mathcal{H}\Phi$\footnote{$\mathcal{H}\Phi$ is
 available at https://github.com/QLMS/HPhi} is used for the exact
 diagonalization calculations.  We thank the computational resources of
 the K computer provided by the RIKEN Advanced Institute for
 Computational Science through the HPCI System Research project
 (hp140136, hp140215, hp150211, hp160201) and the Computational
 Materials Science Initiative (CMSI) supported by Ministry of Education,
 Culture, Sports, Science, and Technology, Japan.  The present work is
 financially supported by JSPS KAKENHI No.~25287097, No.~26287079,
 No.~15K17701, No.~15K17702 and No.~16H06345. This work was funded by ImPACT
 Program of Council for Science, Technology and Innovation (Cabinet
 Office, Government of Japan).
\end{acknowledgements}

\appendix
\section{the {\it ab initio} Hamiltonian of $\mathrm{Na_2IrO_3}$\label{App:Hamiltonian}}
In this appendix, we summarize the derivation of {\it ab initio}
Hamiltonian of $\mathrm{Na_2IrO_3}$ \eqref{eq:Hamiltonian_spin}.

In order to derive the effective spin Hamiltonian from the {\it ab
initio} Hamiltonian for $t_{2g}$ electrons \eqref{eq:Hamiltonian}, we
employ the second order perturbation theory: Here we take the
Hamiltonian of an isolated iridium atom
$\hat{H}_{\mathrm{tri}}+\hat{H}_{\mathrm{SOC}}+\hat{H}_{U}$ as an
unperturbed Hamiltonian and the hopping term $\hat{H}_0$ as a
perturbation\cite{YamajiNKAI2014}. We consider the doubly degenerated
ground state of
$\hat{H}_{\mathrm{tri}}+\hat{H}_{\mathrm{SOC}}+\hat{H}_{U}$ for a
single-site problem as a pseudospin. The exchange couplings among the
pseudospins are derived through the second order perturbation theory by
numerically diagonalizing the Hamiltonian
$\hat{H}_{\mathrm{tri}}+\hat{H}_{\mathrm{SOC}}+\hat{H}_{U}$
\cite{YamajiNKAI2014}.

In Table~\ref{table:exchange_exact}, we show thus obtained exchange
interactions which we used in our {\it ab initio} calculation of the
ground state 
of $\mathrm{Na_2IrO_3}$. Note that for the third-neighbor interaction,
we approximated it as the isotropic Heisenberg interaction where we
neglected off-diagonal interactions ($I_1^{(3rd)}$ and $I_2^{(3rd)}$)
and we averaged the diagonal interactions as $J_3 = (K^{(3rd)} +
2J^{(3rd)})/3$.
\begin{table}
\label{table:exchange_exact} 
\begin{tabular}{c|ccc}
\hline
\hline
\multirow{4}{*}{$J_{X,Y}~(\mathrm{meV})$}&$K'$ & $J'$ & $J''$\\
 \cline{2-4}
 &$-23.9467619$ & $2.0225331$ & $3.2124194$ \\
 \cline{2-4}
 &$I_1'$& $I_2'$&$I_2''$\\
 \cline{2-4}
 &$1.8470590$&$-8.4040133$ &$-3.1148375$\\
\hline
\hline
 \multirow{4}{*}{$J_Z~(\mathrm{meV})$}&$K$&$J$\\
\cline{2-4}
 &$-30.7439117$&$4.4421939$\\
\cline{2-4}
 &$I_1$&$I_2$\\
\cline{2-4}
 &$-0.3777579$&$1.0659292$\\
\hline
\hline
 \multirow{4}{*}{$J_2~(\mathrm{meV})$}&$K^{(2nd)}$&$J^{(2nd)}$\\
\cline{2-4}
 &$-1.2250998$&$-0.8030967$\\
\cline{2-4}
 &$I_1^{(2nd)}$&$I_2^{(2nd)}$\\
\cline{2-4}
 &$0.9901792$&$-1.4245524$\\
\hline
\hline
 \multirow{4}{*}{$J_3~(\mathrm{meV})$}& $K^{(3rd)}$&$J^{(3rd)}$\\
\cline{2-4}
 &$1.7161468$&$1.5996219$\\
\cline{2-4}
 &$I_1^{(3rd)}$&$I_2^{(3rd)}$\\
\cline{2-4}
 &$0.1203473$&$0.0476719$\\
\hline
\hline
\end{tabular}
\caption{Precise exchange interactions of the {\it ab initio}
Kitaev-Heisenberg Hamiltonian at $\Delta = -28$ meV calculated from the
second-order perturbation theory. In the third-neighbor interaction, we
approximated it as the isotropic Heisenberg interaction by neglecting
off-diagonal interactions $I_1^{3rd}$ and $I_2^{3rd}$ and by averaging
diagonal interaction as $J_3 = (K^{(3rd)} + 2J^{(3rd)})/3$}
\end{table}
\section{iPEPS calculation method\label{App:PEPS}}

\begin{figure}
  \includegraphics[width=6cm]{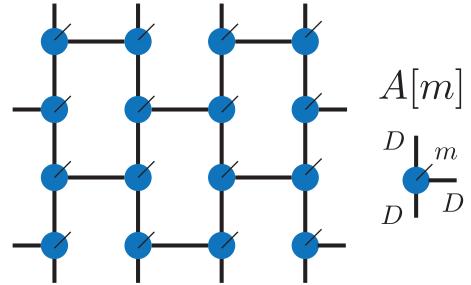} \caption{(Color online)
  Schematic picture of iPEPS tensor network for Honeycomb
  lattice. Tensor $A$ has a physical index and three virtual
  indices. The dimension of the physical index is $m=2$ in the case of
  $S=1/2$ spin system. The dimension of the virtual indices is set to
  $D$, which determines the accuracy of the iPEPS wave function for the
  ground state. The iPSPS becomes exact in the limit $D \rightarrow
  \infty$ and the accuracy is improved systematically with increasing
  $D$.\label{fig:PEPS}}
\end{figure}

In this appendix, we describe methods used in iPEPS calculations. In our
calculation, we first represent the ground-state wave-function
$|\Psi\rangle$ of the model as a tensor product state:
\begin{widetext}
\begin{equation}
 |\Psi\rangle = \sum_{\{m_{\bm{r}_1},m_{\bm{r}_2},\cdots,m_{\bm{r}_i},\cdots\}}\mathrm{Tr}~\prod_{\bm{R}}\left( A_1[m_{1,\bm{R}}]
 |A_2[m_{2,\bm{R}}] \cdots A_i[m_{N,\bm{R}}]\right)  
 |m_{\bm{r}_1}m_{\bm{r_2}}\cdots m_{\bm{r_i}}\cdots \rangle,
\label{eq:iPEPS}
\end{equation}
 \end{widetext}
where $A_i[m]$ is a $4$-rank tensor located at the vertex $i$ of the
 honeycomb lattice with three virtual indices and one physical index,
 $m$, (see Fig.~\ref{fig:PEPS}), and $\mathrm{Tr}$ means the contraction
 over virtual indices.  In order to treat infinite system, we assume
 that the tensors are translationally invariant with a unit cell
 containing $N = L_x \times L_y$ sites, and $m_{\bm{r}} = m_{i,\bm{R}}$
 means the $i$th spin on a unit cell located at $\bm{R}$ (see
 Fig. \ref{fig:unit_cells}). Note that this iPEPS ansatz with a unit
 cell is totally different from a finite $L_x \times L_y$ system with
 the periodic boundary condition. Although the same tensors repeatedly
 appear in the definition of the wave function, (namely, $A_i$ does not
 depend on $\bm{R}$ in Eq.~\eqref{eq:iPEPS}), the spins on equivalent
 but different unit-cell sites can take different values
 $m_{i,\bm{R}}$. Thus, the iPEPS is able to take into account infinitely
 large spin degrees of freedom in contrast to the finite-size algorithm.

Our iPEPS calculations are conducted in two steps. The first step is
optimization of the tensors and the second step is calculation of
physical quantities.

\begin{figure}[h]
  \includegraphics[width=8cm]{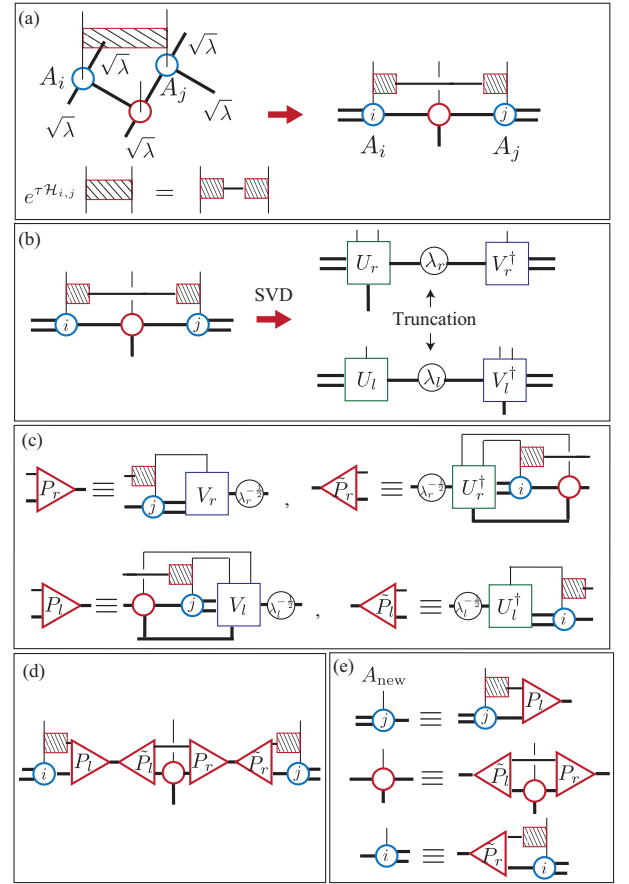} \caption{(Color
  online) Procedure of imaginary-time evolution for the second neighbor
  interaction. Blue and red circles with legs represent tensor $A$
  defined in Eq.~\eqref{eq:iPEPS}, where thick lines are virtual bonds
  and a vertical thin line is the physical bond (see also
  Fig.~\ref{fig:PEPS}). A shaded rectangle represents the imaginary-time
  evolution operator $e^{\tau\mathcal{H}_{i,j}}$. (a) A cluster
  consisting of three tensors ($A_i$, $A_j$ (blue) and a intermediate
  tensor (red)) and an imaginary-time evolution operator with mean field
  like environment ($\sqrt{\lambda}$). Note that because we use tensors
  defined in Eq.~\eqref{eq:iPEPS}, each tensor includes $\sqrt{\lambda}$
  implicitly. Thus, $\sqrt{\lambda}$ is sufficient as the mean-field
  like environment for the simple update. By using a matrix product
  operator representation of $e^{\tau\mathcal{H}_{i,j}}$, we can
  transform the cluster into a one-dimensional chain representation
  shown in the left side. (b) Two types of SVDs which decompose the
  cluster into $2 + 1$ (top) or $1 + 2$ (bottom) segments. Here we
  truncate the tensor by keeping only the largest $D$
  singular values. (c) Construction of two pairs of projectors $(P_r,
  \tilde{P}_r)$ and $(P_l,
  \tilde{P}_l)$. $U_r^\dagger$,$U_l^\dagger$,$V_r$ and $V_l$ are complex
  conjugates of the tensors obtained by SVDs in the step (b). (d) We
  approximate the original cluster by inserting the projectors
  calculated in the step (c). (e) Definition of updated tensors
  $A_{\mathrm{new}}$. \label{fig:PEPS_ITE}}
\end{figure}

In the first step, we optimize tensors using the imaginary-time
evolution by multiplying $e^{\tau\mathcal{H}}$ repeatedly. The
imaginary-time evolution operator is decomposed into a product of
$e^{\tau\mathcal{H}_{ij}}$ with two-body interactions $\mathcal{H}_{ij}$
using Suzuki-Trotter decomposition
\cite{Suzuki1976,Trotter1959}. Typically, we start from $\tau = 0.1/|K|$
and gradually decrease $\tau$ to $\tau= 0.001/|K|$ to reach the ground
state accurately, where $K$ is the largest exchange interaction of the
model. In this imaginary-time evolution, we need a truncation in order
to keep the bond-dimensions of the tensor $A_i[m]$ within a tractable
size. For this truncation we use so called the simple-update method
\cite{JiangWX2008}. In this simple-update method we insert diagonal
matrices $\lambda_{i,j}$ on the bond connecting virtual indices and they
are considered as mean-field like environments at the truncation (see
Fig.~\ref{fig:PEPS_ITE}(a)). For the nearest-neighbor interactions, we
use the singular value decomposition (SVD) and truncate smaller singular
values \cite{JiangWX2008}.

\begin{figure}
  \includegraphics[width=8cm]{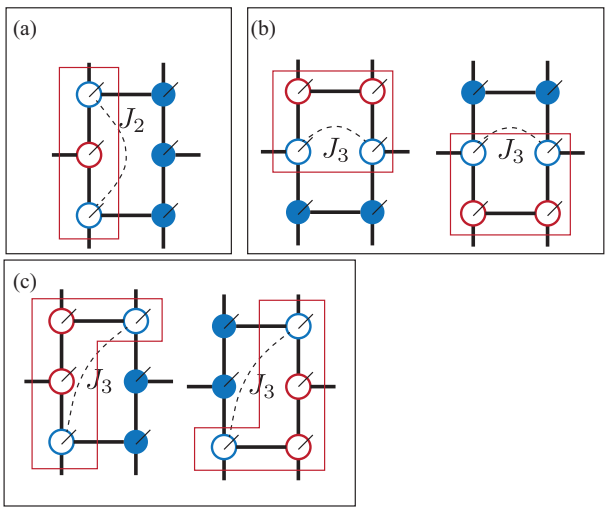} 
  \caption{(Color online) Schematic pictures representing the smallest
  clusters used in imaginary-time evolutions for further neighbor
  interactions. Blue open circles represent tensors interacting through
  a further-neighbor interaction. Red open circles are intermediate
  tensors. The smallest clusters are indicated by red rectangles.  (a)
  In the case of the second-neighbor interaction, the smallest cluster
  is uniquely determined. (b,c) In the case of the third-neighbor
  interactions, there are two types of the smallest cluster.\label{fig:PEPS_cluster}}  
\end{figure}

In the presence of further neighbor interactions, we need to treat at
least three tensors simultaneously because the tensors $A_i$ and $A_j$
interacting through a further-neighbor interaction are not directly
connected in our tensor network. In order to treat further-neighbor
interaction, we connect $A_i$ and $A_j$ through other intermediate
tensor(s) and consider the imaginary-time evolution of the cluster as
shown in Fig.~\ref{fig:PEPS_ITE}(a). In order to construct the clusters,
we use the smallest cluster (the shortest path). If there are more than
one smallest clusters, we decompose the imaginary-time evolution. For
instance, if two smallest clusters exist, we decompose it as
$e^{\tau\mathcal{H}_{ij}}
=e^{\tau\mathcal{H}_{ij}/2}e^{\tau\mathcal{H}_{ij}/2}$, and assign the
different cluster to each of $e^{\tau\mathcal{H}_{ij}/2}$. In the case
of the {\it ab initio} Kitaev-Heisenberg Hamiltonian, the smallest
cluster is unique for $J_2$ interaction (see
Fig.~\ref{fig:PEPS_cluster}(a)), while there are two types of the
smallest cluster for $J_3$ interactions (see
Figs.~\ref{fig:PEPS_cluster}(b) and (c)).

In order to decompose the cluster, we use symmetric decomposition by
constructing projectors, which is slightly different from the
decomposition using successive SVDs \cite{CorbozJV2010}.

In the case of a cluster consisting of three tensors, first we construct a
cluster by combining three $A$ tensors, mean-field like environments
$\sqrt{\lambda}$, and the imaginary-time evolution operator as shown in
Fig.~\ref{fig:PEPS_ITE} (a). Note that we can see the cluster as a
one-dimensional chain shown in the left part of
Fig.~\ref{fig:PEPS_ITE}(a), where for the sake of flowing calculations
we introduce the matrix product operator (MPO) representation of the
imaginary-time evolution operator.

Next, we perform two SVDs which decompose the cluster into $2 + 1$ and $1 +2$
segments (see Fig.~\ref{fig:PEPS_ITE}(b)). From these SVDs, we obtain
two sets of tensors and singular values, $(U_r, \lambda_r, V_r^\dagger)$
and $(U_l, \lambda_l, V_l^\dagger)$.

Then, we construct two pairs of projectors $(P_r$, $\tilde{P}_r)$ and
$(P_l$, $\tilde{P}_l)$ by using tensors obtained by the previous SVDs as
shown in Fig.~\ref{fig:PEPS_ITE}(c). Note that these projectors satisfy
the relation $\tilde{P}P = \text{identity}$.

Finally, we insert projectors into the cluster
(Fig.~\ref{fig:PEPS_ITE}(d)) and decompose it into three parts to
obtain updated tensors. The updated tensors are defined as
Fig.~\ref{fig:PEPS_ITE} (e).

For larger clusters we can use the same method by considering several SVDs
and creating projectors. In the actual calculation, we perform QR
decomposition of tensors $A_i$ and $A_j$ before applying imaginary-time
evolution operator in order to reduce the computational
cost \cite{WangPV2011,DepenbrockP2013}.

In the second step, we calculate expectation values from the obtained
wave function. In this step, we use the approximate contraction based on
the corner transfer matrix (CTM) method
\cite{Baxter1968,Baxter1978,Baxter_book,NishinoO1998,OrusV2009,CorbozWV2011,CorbozRT2014}. In
order to treat the several unit-cell shapes, we use the directional CTM
renormalization group \cite{OrusV2009}, with the generalization to
arbitrary unit-cell sizes \cite{CorbozWV2011,CorbozRT2014}. As the bond
dimension $\chi$ of the CTMs, we typically use $\chi = D^2$ because
further increase of $\chi$ beyond $D^2$ did not change expectation
values largely in the case of magnetically ordered phase we observed.

\section{iPEPS calculation for the nearest neighbor
model\label{App:PEPS_NN}}

In this appendix, to compare with the full {\it ab initio} studies and
to gain further insights, we briefly show the analysis on a simplified
nearest-neighbor {\it ab initio} Hamiltonian for $\mathrm{Na_2IrO_3}$
where we neglect the second and the third neighbor interactions and
consider only the nearest-neighbor interactions. For this analysis, we
use tensor network methods with the iPEPS ansatz. In the case of the
nearest-neighbor interaction, we can easily apply the so called ``full
update'' method\cite{JordanOVVC2008,OrusV2009}, which is expected to be
more accurate than the simple update used in the analysis of the main
part.  By considering the nearest-neighbor interaction only, we can
compare the results obtained by the simple update and the full update,
and estimate the reliability of simple update method.
\begin{figure}
  \includegraphics[width=8cm]{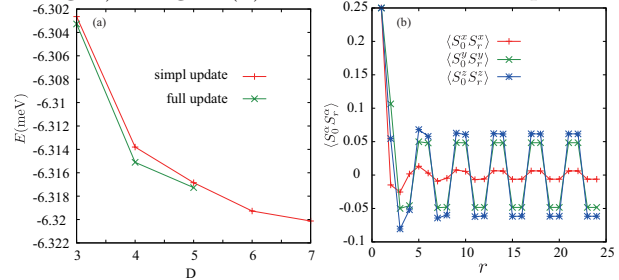} \caption{(Color
  online) (a) Bond-dimension ($D$) dependence of the energy of the
  nearest-neighbor {\it ab initio} Hamiltonian calculated by the iPEPS
  optimized by using the simple update and the full update with the $L_x
  \times L_y = 2\times 4$ unit cell. (b) Spin-spin correlation function
  of $S^x, S^y$, and $S^z$ along the $y$-direction obtained by iPEPS
  with the simple update ($D=7$).}  \label{fig:NN_enemag}
\end{figure}

In Fig.~\ref{fig:NN_enemag}(a), we show the energy of the nearest-neighbor
{\it ab initio} Hamiltonian as a function of the bond dimension
$D$. Although the energies obtained by the full update method is
slightly lower than those of the simple update at the same $D$, the
difference is quite small compared with the decrease in the energy with
increasing the bond dimension. Thus, the simple update seems to be
sufficiently reliable for the present {\it ab initio} Hamiltonian whose
ground state is expected to be a magnetically ordered state.

In order to further investigate the nature of the ground state, we show
the spin correlation along the $y$-direction (see
Fig.~\ref{fig:unit_cells}) in Fig.~\ref{fig:NN_enemag}(b). One can see
that the spin correlation shows four-site periodicity, which is totally
different from the zigzag($Z$) structure observed in the ground state of
{\it ab initio} Hamiltonian including the second and the further
neighbor interactions. Indeed, the ground state of the nearest-neighbor
{\it ab initio} Hamiltonian is an $8$-site state different from the
zigzag($Z$) state. Thus, in order to obtain the experimentally observed
zigzag($Z$) state, the further-neighbor interactions are crucially
important.

Note that the iPEPS has been able to describe the Kitaev spin liquid
state by using the full update optimization if it is applied to the
Kitaev-Heisenberg model \cite{IreguiCT2014}. Therefore, the method is
capable of describing the Kitaev spin liquid in general. Nevertheless,
we did not obtain the spin liquid state for a more realistic Hamiltonian
with only the nearest-neighbor interaction truncated from the {\it ab
initio} Hamiltonian even if we used the full update. Furthermore, the
further-neighbor interactions stabilize the magnetically ordered state
rather than the Kitaev spin liquid, as we see in the ground state of the
full {\it ab initio} Hamiltonian, which includes the second and the
third neighbor interactions. Therefore, our result shows that the
magnetic order rather than the spin liquid is robust around the {\it ab
initio} parameter values, consistently with the experimental results.

%
%

\bibliography{Kitaev_Heisenberg}

\begin{thebibliography}{43}%
\makeatletter
\providecommand \@ifxundefined [1]{%
 \@ifx{#1\undefined}
}%
\providecommand \@ifnum [1]{%
 \ifnum #1\expandafter \@firstoftwo
 \else \expandafter \@secondoftwo
 \fi
}%
\providecommand \@ifx [1]{%
 \ifx #1\expandafter \@firstoftwo
 \else \expandafter \@secondoftwo
 \fi
}%
\providecommand \natexlab [1]{#1}%
\providecommand \enquote  [1]{``#1''}%
\providecommand \bibnamefont  [1]{#1}%
\providecommand \bibfnamefont [1]{#1}%
\providecommand \citenamefont [1]{#1}%
\providecommand \href@noop [0]{\@secondoftwo}%
\providecommand \href [0]{\begingroup \@sanitize@url \@href}%
\providecommand \@href[1]{\@@startlink{#1}\@@href}%
\providecommand \@@href[1]{\endgroup#1\@@endlink}%
\providecommand \@sanitize@url [0]{\catcode `\\12\catcode `\$12\catcode
  `\&12\catcode `\#12\catcode `\^12\catcode `\_12\catcode `\%12\relax}%
\providecommand \@@startlink[1]{}%
\providecommand \@@endlink[0]{}%
\providecommand \url  [0]{\begingroup\@sanitize@url \@url }%
\providecommand \@url [1]{\endgroup\@href {#1}{\urlprefix }}%
\providecommand \urlprefix  [0]{URL }%
\providecommand \Eprint [0]{\href }%
\providecommand \doibase [0]{http://dx.doi.org/}%
\providecommand \selectlanguage [0]{\@gobble}%
\providecommand \bibinfo  [0]{\@secondoftwo}%
\providecommand \bibfield  [0]{\@secondoftwo}%
\providecommand \translation [1]{[#1]}%
\providecommand \BibitemOpen [0]{}%
\providecommand \bibitemStop [0]{}%
\providecommand \bibitemNoStop [0]{.\EOS\space}%
\providecommand \EOS [0]{\spacefactor3000\relax}%
\providecommand \BibitemShut  [1]{\csname bibitem#1\endcsname}%
\let\auto@bib@innerbib\@empty
\bibitem [{\citenamefont {Jackeli}\ and\ \citenamefont
  {Khaliullin}(2009)}]{JackeliK2009}%
  \BibitemOpen
  \bibfield  {author} {\bibinfo {author} {\bibfnamefont {G.}~\bibnamefont
  {Jackeli}}\ and\ \bibinfo {author} {\bibfnamefont {G.}~\bibnamefont
  {Khaliullin}},\ }\href {\doibase 10.1103/PhysRevLett.102.017205} {\bibfield
  {journal} {\bibinfo  {journal} {Phys. Rev. Lett.}\ }\textbf {\bibinfo
  {volume} {102}},\ \bibinfo {pages} {017205} (\bibinfo {year}
  {2009})}\BibitemShut {NoStop}%
\bibitem [{\citenamefont {Chaloupka}\ \emph {et~al.}(2010)\citenamefont
  {Chaloupka}, \citenamefont {Jackeli},\ and\ \citenamefont
  {Khaliullin}}]{ChaloupkaJK2010}%
  \BibitemOpen
  \bibfield  {author} {\bibinfo {author} {\bibfnamefont {J.~c.~v.}\
  \bibnamefont {Chaloupka}}, \bibinfo {author} {\bibfnamefont {G.}~\bibnamefont
  {Jackeli}}, \ and\ \bibinfo {author} {\bibfnamefont {G.}~\bibnamefont
  {Khaliullin}},\ }\href {\doibase 10.1103/PhysRevLett.105.027204} {\bibfield
  {journal} {\bibinfo  {journal} {Phys. Rev. Lett.}\ }\textbf {\bibinfo
  {volume} {105}},\ \bibinfo {pages} {027204} (\bibinfo {year}
  {2010})}\BibitemShut {NoStop}%
\bibitem [{\citenamefont {Wan}\ \emph {et~al.}(2011)\citenamefont {Wan},
  \citenamefont {Turner}, \citenamefont {Vishwanath},\ and\ \citenamefont
  {Savrasov}}]{WanTVS2011}%
  \BibitemOpen
  \bibfield  {author} {\bibinfo {author} {\bibfnamefont {X.}~\bibnamefont
  {Wan}}, \bibinfo {author} {\bibfnamefont {A.~M.}\ \bibnamefont {Turner}},
  \bibinfo {author} {\bibfnamefont {A.}~\bibnamefont {Vishwanath}}, \ and\
  \bibinfo {author} {\bibfnamefont {S.~Y.}\ \bibnamefont {Savrasov}},\ }\href
  {\doibase 10.1103/PhysRevB.83.205101} {\bibfield  {journal} {\bibinfo
  {journal} {Phys. Rev. B}\ }\textbf {\bibinfo {volume} {83}},\ \bibinfo
  {pages} {205101} (\bibinfo {year} {2011})}\BibitemShut {NoStop}%
\bibitem [{\citenamefont {Witczak-Krempa}\ \emph {et~al.}(2014)\citenamefont
  {Witczak-Krempa}, \citenamefont {Chen}, \citenamefont {Kim},\ and\
  \citenamefont {Balents}}]{WitczakCKB2014}%
  \BibitemOpen
  \bibfield  {author} {\bibinfo {author} {\bibfnamefont {W.}~\bibnamefont
  {Witczak-Krempa}}, \bibinfo {author} {\bibfnamefont {G.}~\bibnamefont
  {Chen}}, \bibinfo {author} {\bibfnamefont {Y.~B.}\ \bibnamefont {Kim}}, \
  and\ \bibinfo {author} {\bibfnamefont {L.}~\bibnamefont {Balents}},\ }\href
  {\doibase 10.1146/annurev-conmatphys-020911-125138} {\bibfield  {journal}
  {\bibinfo  {journal} {Annual Review of Condensed Matter Physics}\ }\textbf
  {\bibinfo {volume} {5}},\ \bibinfo {pages} {57} (\bibinfo {year}
  {2014})}\BibitemShut {NoStop}%
\bibitem [{\citenamefont {Liu}\ \emph {et~al.}(2011)\citenamefont {Liu},
  \citenamefont {Berlijn}, \citenamefont {Yin}, \citenamefont {Ku},
  \citenamefont {Tsvelik}, \citenamefont {Kim}, \citenamefont {Gretarsson},
  \citenamefont {Singh}, \citenamefont {Gegenwart},\ and\ \citenamefont
  {Hill}}]{Liu2011}%
  \BibitemOpen
  \bibfield  {author} {\bibinfo {author} {\bibfnamefont {X.}~\bibnamefont
  {Liu}}, \bibinfo {author} {\bibfnamefont {T.}~\bibnamefont {Berlijn}},
  \bibinfo {author} {\bibfnamefont {W.-G.}\ \bibnamefont {Yin}}, \bibinfo
  {author} {\bibfnamefont {W.}~\bibnamefont {Ku}}, \bibinfo {author}
  {\bibfnamefont {A.}~\bibnamefont {Tsvelik}}, \bibinfo {author} {\bibfnamefont
  {Y.-J.}\ \bibnamefont {Kim}}, \bibinfo {author} {\bibfnamefont
  {H.}~\bibnamefont {Gretarsson}}, \bibinfo {author} {\bibfnamefont
  {Y.}~\bibnamefont {Singh}}, \bibinfo {author} {\bibfnamefont
  {P.}~\bibnamefont {Gegenwart}}, \ and\ \bibinfo {author} {\bibfnamefont
  {J.~P.}\ \bibnamefont {Hill}},\ }\href {\doibase 10.1103/PhysRevB.83.220403}
  {\bibfield  {journal} {\bibinfo  {journal} {Phys. Rev. B}\ }\textbf {\bibinfo
  {volume} {83}},\ \bibinfo {pages} {220403} (\bibinfo {year}
  {2011})}\BibitemShut {NoStop}%
\bibitem [{\citenamefont {Choi}\ \emph {et~al.}(2012)\citenamefont {Choi},
  \citenamefont {Coldea}, \citenamefont {Kolmogorov}, \citenamefont
  {Lancaster}, \citenamefont {Mazin}, \citenamefont {Blundell}, \citenamefont
  {Radaelli}, \citenamefont {Singh}, \citenamefont {Gegenwart}, \citenamefont
  {Choi}, \citenamefont {Cheong}, \citenamefont {Baker}, \citenamefont
  {Stock},\ and\ \citenamefont {Taylor}}]{Choi2012}%
  \BibitemOpen
  \bibfield  {author} {\bibinfo {author} {\bibfnamefont {S.~K.}\ \bibnamefont
  {Choi}}, \bibinfo {author} {\bibfnamefont {R.}~\bibnamefont {Coldea}},
  \bibinfo {author} {\bibfnamefont {A.~N.}\ \bibnamefont {Kolmogorov}},
  \bibinfo {author} {\bibfnamefont {T.}~\bibnamefont {Lancaster}}, \bibinfo
  {author} {\bibfnamefont {I.~I.}\ \bibnamefont {Mazin}}, \bibinfo {author}
  {\bibfnamefont {S.~J.}\ \bibnamefont {Blundell}}, \bibinfo {author}
  {\bibfnamefont {P.~G.}\ \bibnamefont {Radaelli}}, \bibinfo {author}
  {\bibfnamefont {Y.}~\bibnamefont {Singh}}, \bibinfo {author} {\bibfnamefont
  {P.}~\bibnamefont {Gegenwart}}, \bibinfo {author} {\bibfnamefont {K.~R.}\
  \bibnamefont {Choi}}, \bibinfo {author} {\bibfnamefont {S.-W.}\ \bibnamefont
  {Cheong}}, \bibinfo {author} {\bibfnamefont {P.~J.}\ \bibnamefont {Baker}},
  \bibinfo {author} {\bibfnamefont {C.}~\bibnamefont {Stock}}, \ and\ \bibinfo
  {author} {\bibfnamefont {J.}~\bibnamefont {Taylor}},\ }\href {\doibase
  10.1103/PhysRevLett.108.127204} {\bibfield  {journal} {\bibinfo  {journal}
  {Phys. Rev. Lett.}\ }\textbf {\bibinfo {volume} {108}},\ \bibinfo {pages}
  {127204} (\bibinfo {year} {2012})}\BibitemShut {NoStop}%
\bibitem [{\citenamefont {Ye}\ \emph {et~al.}(2012)\citenamefont {Ye},
  \citenamefont {Chi}, \citenamefont {Cao}, \citenamefont {Chakoumakos},
  \citenamefont {Fernandez-Baca}, \citenamefont {Custelcean}, \citenamefont
  {Qi}, \citenamefont {Korneta},\ and\ \citenamefont {Cao}}]{Ye2012}%
  \BibitemOpen
  \bibfield  {author} {\bibinfo {author} {\bibfnamefont {F.}~\bibnamefont
  {Ye}}, \bibinfo {author} {\bibfnamefont {S.}~\bibnamefont {Chi}}, \bibinfo
  {author} {\bibfnamefont {H.}~\bibnamefont {Cao}}, \bibinfo {author}
  {\bibfnamefont {B.~C.}\ \bibnamefont {Chakoumakos}}, \bibinfo {author}
  {\bibfnamefont {J.~A.}\ \bibnamefont {Fernandez-Baca}}, \bibinfo {author}
  {\bibfnamefont {R.}~\bibnamefont {Custelcean}}, \bibinfo {author}
  {\bibfnamefont {T.~F.}\ \bibnamefont {Qi}}, \bibinfo {author} {\bibfnamefont
  {O.~B.}\ \bibnamefont {Korneta}}, \ and\ \bibinfo {author} {\bibfnamefont
  {G.}~\bibnamefont {Cao}},\ }\href {\doibase 10.1103/PhysRevB.85.180403}
  {\bibfield  {journal} {\bibinfo  {journal} {Phys. Rev. B}\ }\textbf {\bibinfo
  {volume} {85}},\ \bibinfo {pages} {180403} (\bibinfo {year}
  {2012})}\BibitemShut {NoStop}%
\bibitem [{\citenamefont {Lovesey}\ and\ \citenamefont
  {Dobrynin}(2012)}]{LoveseyD2012}%
  \BibitemOpen
  \bibfield  {author} {\bibinfo {author} {\bibfnamefont {S.~W.}\ \bibnamefont
  {Lovesey}}\ and\ \bibinfo {author} {\bibfnamefont {A.~N.}\ \bibnamefont
  {Dobrynin}},\ }\href {http://stacks.iop.org/0953-8984/24/i=38/a=382201}
  {\bibfield  {journal} {\bibinfo  {journal} {Journal of Physics: Condensed
  Matter}\ }\textbf {\bibinfo {volume} {24}},\ \bibinfo {pages} {382201}
  (\bibinfo {year} {2012})}\BibitemShut {NoStop}%
\bibitem [{\citenamefont {Alpichshev}\ \emph {et~al.}(2015)\citenamefont
  {Alpichshev}, \citenamefont {Mahmood}, \citenamefont {Cao},\ and\
  \citenamefont {Gedik}}]{AlpichshevMCG2015}%
  \BibitemOpen
  \bibfield  {author} {\bibinfo {author} {\bibfnamefont {Z.}~\bibnamefont
  {Alpichshev}}, \bibinfo {author} {\bibfnamefont {F.}~\bibnamefont {Mahmood}},
  \bibinfo {author} {\bibfnamefont {G.}~\bibnamefont {Cao}}, \ and\ \bibinfo
  {author} {\bibfnamefont {N.}~\bibnamefont {Gedik}},\ }\href {\doibase
  10.1103/PhysRevLett.114.017203} {\bibfield  {journal} {\bibinfo  {journal}
  {Phys. Rev. Lett.}\ }\textbf {\bibinfo {volume} {114}},\ \bibinfo {pages}
  {017203} (\bibinfo {year} {2015})}\BibitemShut {NoStop}%
\bibitem [{\citenamefont {Singh}\ \emph {et~al.}(2012)\citenamefont {Singh},
  \citenamefont {Manni}, \citenamefont {Reuther}, \citenamefont {Berlijn},
  \citenamefont {Thomale}, \citenamefont {Ku}, \citenamefont {Trebst},\ and\
  \citenamefont {Gegenwart}}]{SinghMRBTKTG2012}%
  \BibitemOpen
  \bibfield  {author} {\bibinfo {author} {\bibfnamefont {Y.}~\bibnamefont
  {Singh}}, \bibinfo {author} {\bibfnamefont {S.}~\bibnamefont {Manni}},
  \bibinfo {author} {\bibfnamefont {J.}~\bibnamefont {Reuther}}, \bibinfo
  {author} {\bibfnamefont {T.}~\bibnamefont {Berlijn}}, \bibinfo {author}
  {\bibfnamefont {R.}~\bibnamefont {Thomale}}, \bibinfo {author} {\bibfnamefont
  {W.}~\bibnamefont {Ku}}, \bibinfo {author} {\bibfnamefont {S.}~\bibnamefont
  {Trebst}}, \ and\ \bibinfo {author} {\bibfnamefont {P.}~\bibnamefont
  {Gegenwart}},\ }\href {\doibase 10.1103/PhysRevLett.108.127203} {\bibfield
  {journal} {\bibinfo  {journal} {Phys. Rev. Lett.}\ }\textbf {\bibinfo
  {volume} {108}},\ \bibinfo {pages} {127203} (\bibinfo {year}
  {2012})}\BibitemShut {NoStop}%
\bibitem [{\citenamefont {Kitaev}(2006)}]{Kitaev2006}%
  \BibitemOpen
  \bibfield  {author} {\bibinfo {author} {\bibfnamefont {A.}~\bibnamefont
  {Kitaev}},\ }\href {\doibase http://dx.doi.org/10.1016/j.aop.2005.10.005}
  {\bibfield  {journal} {\bibinfo  {journal} {Annals of Physics}\ }\textbf
  {\bibinfo {volume} {321}},\ \bibinfo {pages} {2 } (\bibinfo {year} {2006})},\
  \bibinfo {note} {january Special Issue}\BibitemShut {NoStop}%
\bibitem [{\citenamefont {Kimchi}\ and\ \citenamefont
  {You}(2011)}]{KimuchiY2011}%
  \BibitemOpen
  \bibfield  {author} {\bibinfo {author} {\bibfnamefont {I.}~\bibnamefont
  {Kimchi}}\ and\ \bibinfo {author} {\bibfnamefont {Y.-Z.}\ \bibnamefont
  {You}},\ }\href {\doibase 10.1103/PhysRevB.84.180407} {\bibfield  {journal}
  {\bibinfo  {journal} {Phys. Rev. B}\ }\textbf {\bibinfo {volume} {84}},\
  \bibinfo {pages} {180407} (\bibinfo {year} {2011})}\BibitemShut {NoStop}%
\bibitem [{\citenamefont {Chaloupka}\ \emph {et~al.}(2013)\citenamefont
  {Chaloupka}, \citenamefont {Jackeli},\ and\ \citenamefont
  {Khaliullin}}]{ChaloupkaJK2013}%
  \BibitemOpen
  \bibfield  {author} {\bibinfo {author} {\bibfnamefont {J.~c.~v.}\
  \bibnamefont {Chaloupka}}, \bibinfo {author} {\bibfnamefont {G.}~\bibnamefont
  {Jackeli}}, \ and\ \bibinfo {author} {\bibfnamefont {G.}~\bibnamefont
  {Khaliullin}},\ }\href {\doibase 10.1103/PhysRevLett.110.097204} {\bibfield
  {journal} {\bibinfo  {journal} {Phys. Rev. Lett.}\ }\textbf {\bibinfo
  {volume} {110}},\ \bibinfo {pages} {097204} (\bibinfo {year}
  {2013})}\BibitemShut {NoStop}%
\bibitem [{\citenamefont {Sizyuk}\ \emph {et~al.}(2014)\citenamefont {Sizyuk},
  \citenamefont {Price}, \citenamefont {W\"olfle},\ and\ \citenamefont
  {Perkins}}]{SizyukPWP2014}%
  \BibitemOpen
  \bibfield  {author} {\bibinfo {author} {\bibfnamefont {Y.}~\bibnamefont
  {Sizyuk}}, \bibinfo {author} {\bibfnamefont {C.}~\bibnamefont {Price}},
  \bibinfo {author} {\bibfnamefont {P.}~\bibnamefont {W\"olfle}}, \ and\
  \bibinfo {author} {\bibfnamefont {N.~B.}\ \bibnamefont {Perkins}},\ }\href
  {\doibase 10.1103/PhysRevB.90.155126} {\bibfield  {journal} {\bibinfo
  {journal} {Phys. Rev. B}\ }\textbf {\bibinfo {volume} {90}},\ \bibinfo
  {pages} {155126} (\bibinfo {year} {2014})}\BibitemShut {NoStop}%
\bibitem [{\citenamefont {Bhattacharjee}\ \emph {et~al.}(2012)\citenamefont
  {Bhattacharjee}, \citenamefont {Lee},\ and\ \citenamefont
  {Kim}}]{BhattacharjeeLK2012}%
  \BibitemOpen
  \bibfield  {author} {\bibinfo {author} {\bibfnamefont {S.}~\bibnamefont
  {Bhattacharjee}}, \bibinfo {author} {\bibfnamefont {S.-S.}\ \bibnamefont
  {Lee}}, \ and\ \bibinfo {author} {\bibfnamefont {Y.~B.}\ \bibnamefont
  {Kim}},\ }\href {http://stacks.iop.org/1367-2630/14/i=7/a=073015} {\bibfield
  {journal} {\bibinfo  {journal} {New Journal of Physics}\ }\textbf {\bibinfo
  {volume} {14}},\ \bibinfo {pages} {073015} (\bibinfo {year}
  {2012})}\BibitemShut {NoStop}%
\bibitem [{\citenamefont {Rau}\ \emph {et~al.}(2014)\citenamefont {Rau},
  \citenamefont {Lee},\ and\ \citenamefont {Kee}}]{RauLK2014}%
  \BibitemOpen
  \bibfield  {author} {\bibinfo {author} {\bibfnamefont {J.~G.}\ \bibnamefont
  {Rau}}, \bibinfo {author} {\bibfnamefont {E.~K.-H.}\ \bibnamefont {Lee}}, \
  and\ \bibinfo {author} {\bibfnamefont {H.-Y.}\ \bibnamefont {Kee}},\ }\href
  {\doibase 10.1103/PhysRevLett.112.077204} {\bibfield  {journal} {\bibinfo
  {journal} {Phys. Rev. Lett.}\ }\textbf {\bibinfo {volume} {112}},\ \bibinfo
  {pages} {077204} (\bibinfo {year} {2014})}\BibitemShut {NoStop}%
\bibitem [{\citenamefont {Katukuri}\ \emph {et~al.}(2014)\citenamefont
  {Katukuri}, \citenamefont {Nishimoto}, \citenamefont {Yushankhai},
  \citenamefont {Stoyanova}, \citenamefont {Kandpal}, \citenamefont {Choi},
  \citenamefont {Coldea}, \citenamefont {Rousochatzakis}, \citenamefont
  {Hozoi},\ and\ \citenamefont {van~den Brink}}]{KatukuriNYSK2014}%
  \BibitemOpen
  \bibfield  {author} {\bibinfo {author} {\bibfnamefont {V.~M.}\ \bibnamefont
  {Katukuri}}, \bibinfo {author} {\bibfnamefont {S.}~\bibnamefont {Nishimoto}},
  \bibinfo {author} {\bibfnamefont {V.}~\bibnamefont {Yushankhai}}, \bibinfo
  {author} {\bibfnamefont {A.}~\bibnamefont {Stoyanova}}, \bibinfo {author}
  {\bibfnamefont {H.}~\bibnamefont {Kandpal}}, \bibinfo {author} {\bibfnamefont
  {S.}~\bibnamefont {Choi}}, \bibinfo {author} {\bibfnamefont {R.}~\bibnamefont
  {Coldea}}, \bibinfo {author} {\bibfnamefont {I.}~\bibnamefont
  {Rousochatzakis}}, \bibinfo {author} {\bibfnamefont {L.}~\bibnamefont
  {Hozoi}}, \ and\ \bibinfo {author} {\bibfnamefont {J.}~\bibnamefont {van~den
  Brink}},\ }\href {http://stacks.iop.org/1367-2630/16/i=1/a=013056} {\bibfield
   {journal} {\bibinfo  {journal} {New Journal of Physics}\ }\textbf {\bibinfo
  {volume} {16}},\ \bibinfo {pages} {013056} (\bibinfo {year}
  {2014})}\BibitemShut {NoStop}%
\bibitem [{\citenamefont {Yamaji}\ \emph {et~al.}(2014)\citenamefont {Yamaji},
  \citenamefont {Nomura}, \citenamefont {Kurita}, \citenamefont {Arita},\ and\
  \citenamefont {Imada}}]{YamajiNKAI2014}%
  \BibitemOpen
  \bibfield  {author} {\bibinfo {author} {\bibfnamefont {Y.}~\bibnamefont
  {Yamaji}}, \bibinfo {author} {\bibfnamefont {Y.}~\bibnamefont {Nomura}},
  \bibinfo {author} {\bibfnamefont {M.}~\bibnamefont {Kurita}}, \bibinfo
  {author} {\bibfnamefont {R.}~\bibnamefont {Arita}}, \ and\ \bibinfo {author}
  {\bibfnamefont {M.}~\bibnamefont {Imada}},\ }\href {\doibase
  10.1103/PhysRevLett.113.107201} {\bibfield  {journal} {\bibinfo  {journal}
  {Phys. Rev. Lett.}\ }\textbf {\bibinfo {volume} {113}},\ \bibinfo {pages}
  {107201} (\bibinfo {year} {2014})}\BibitemShut {NoStop}%
\bibitem [{\citenamefont {{Rau}}\ and\ \citenamefont
  {{Kee}}(2014)}]{RauK2014arXiv}%
  \BibitemOpen
  \bibfield  {author} {\bibinfo {author} {\bibfnamefont {J.~G.}\ \bibnamefont
  {{Rau}}}\ and\ \bibinfo {author} {\bibfnamefont {H.-Y.}\ \bibnamefont
  {{Kee}}},\ }\href@noop {} {\bibfield  {journal} {\bibinfo  {journal}
  {arXiv:1408.4811}\ } (\bibinfo {year} {2014})}\BibitemShut {NoStop}%
\bibitem [{\citenamefont {Yamaji}\ \emph {et~al.}(2016)\citenamefont {Yamaji},
  \citenamefont {Suzuki}, \citenamefont {Yamada}, \citenamefont {Suga},
  \citenamefont {Kawashima},\ and\ \citenamefont {Imada}}]{YamajiSYSKI2016}%
  \BibitemOpen
  \bibfield  {author} {\bibinfo {author} {\bibfnamefont {Y.}~\bibnamefont
  {Yamaji}}, \bibinfo {author} {\bibfnamefont {T.}~\bibnamefont {Suzuki}},
  \bibinfo {author} {\bibfnamefont {T.}~\bibnamefont {Yamada}}, \bibinfo
  {author} {\bibfnamefont {S.-i.}\ \bibnamefont {Suga}}, \bibinfo {author}
  {\bibfnamefont {N.}~\bibnamefont {Kawashima}}, \ and\ \bibinfo {author}
  {\bibfnamefont {M.}~\bibnamefont {Imada}},\ }\href {\doibase
  10.1103/PhysRevB.93.174425} {\bibfield  {journal} {\bibinfo  {journal} {Phys.
  Rev. B}\ }\textbf {\bibinfo {volume} {93}},\ \bibinfo {pages} {174425}
  (\bibinfo {year} {2016})}\BibitemShut {NoStop}%
\bibitem [{\citenamefont {Osorio~Iregui}\ \emph {et~al.}(2014)\citenamefont
  {Osorio~Iregui}, \citenamefont {Corboz},\ and\ \citenamefont
  {Troyer}}]{IreguiCT2014}%
  \BibitemOpen
  \bibfield  {author} {\bibinfo {author} {\bibfnamefont {J.}~\bibnamefont
  {Osorio~Iregui}}, \bibinfo {author} {\bibfnamefont {P.}~\bibnamefont
  {Corboz}}, \ and\ \bibinfo {author} {\bibfnamefont {M.}~\bibnamefont
  {Troyer}},\ }\href {\doibase 10.1103/PhysRevB.90.195102} {\bibfield
  {journal} {\bibinfo  {journal} {Phys. Rev. B}\ }\textbf {\bibinfo {volume}
  {90}},\ \bibinfo {pages} {195102} (\bibinfo {year} {2014})}\BibitemShut
  {NoStop}%
\bibitem [{\citenamefont {{Villain, J.}}\ \emph {et~al.}(1980)\citenamefont
  {{Villain, J.}}, \citenamefont {{Bidaux, R.}}, \citenamefont {{Carton,
  J.-P.}},\ and\ \citenamefont {{Conte, R.}}}]{VillainBCC1980}%
  \BibitemOpen
  \bibfield  {author} {\bibinfo {author} {\bibnamefont {{Villain, J.}}},
  \bibinfo {author} {\bibnamefont {{Bidaux, R.}}}, \bibinfo {author}
  {\bibnamefont {{Carton, J.-P.}}}, \ and\ \bibinfo {author} {\bibnamefont
  {{Conte, R.}}},\ }\href {\doibase 10.1051/jphys:0198000410110126300}
  {\bibfield  {journal} {\bibinfo  {journal} {J. Phys. France}\ }\textbf
  {\bibinfo {volume} {41}},\ \bibinfo {pages} {1263} (\bibinfo {year}
  {1980})}\BibitemShut {NoStop}%
\bibitem [{\citenamefont {Shinjo}\ \emph {et~al.}(2015)\citenamefont {Shinjo},
  \citenamefont {Sota},\ and\ \citenamefont {Tohyama}}]{ShinjoST2015}%
  \BibitemOpen
  \bibfield  {author} {\bibinfo {author} {\bibfnamefont {K.}~\bibnamefont
  {Shinjo}}, \bibinfo {author} {\bibfnamefont {S.}~\bibnamefont {Sota}}, \ and\
  \bibinfo {author} {\bibfnamefont {T.}~\bibnamefont {Tohyama}},\ }\href
  {\doibase 10.1103/PhysRevB.91.054401} {\bibfield  {journal} {\bibinfo
  {journal} {Phys. Rev. B}\ }\textbf {\bibinfo {volume} {91}},\ \bibinfo
  {pages} {054401} (\bibinfo {year} {2015})}\BibitemShut {NoStop}%
\bibitem [{\citenamefont {Verstraete}\ and\ \citenamefont
  {Cirac}(2004)}]{VerstraeteC2004}%
  \BibitemOpen
  \bibfield  {author} {\bibinfo {author} {\bibfnamefont {F.}~\bibnamefont
  {Verstraete}}\ and\ \bibinfo {author} {\bibfnamefont {J.~I.}\ \bibnamefont
  {Cirac}},\ }\href {\doibase 10.1103/PhysRevA.70.060302} {\bibfield  {journal}
  {\bibinfo  {journal} {Phys. Rev. A}\ }\textbf {\bibinfo {volume} {70}},\
  \bibinfo {pages} {060302} (\bibinfo {year} {2004})}\BibitemShut {NoStop}%
\bibitem [{\citenamefont {{Verstraete}}\ and\ \citenamefont
  {{Cirac}}(2004)}]{VerstraeteC2004arXiv}%
  \BibitemOpen
  \bibfield  {author} {\bibinfo {author} {\bibfnamefont {F.}~\bibnamefont
  {{Verstraete}}}\ and\ \bibinfo {author} {\bibfnamefont {J.~I.}\ \bibnamefont
  {{Cirac}}},\ }\href@noop {} {\bibfield  {journal} {\bibinfo  {journal}
  {arXiv:cond-mat/0407066}\ } (\bibinfo {year} {2004})}\BibitemShut {NoStop}%
\bibitem [{\citenamefont {Jordan}\ \emph {et~al.}(2008)\citenamefont {Jordan},
  \citenamefont {Or\'us}, \citenamefont {Vidal}, \citenamefont {Verstraete},\
  and\ \citenamefont {Cirac}}]{JordanOVVC2008}%
  \BibitemOpen
  \bibfield  {author} {\bibinfo {author} {\bibfnamefont {J.}~\bibnamefont
  {Jordan}}, \bibinfo {author} {\bibfnamefont {R.}~\bibnamefont {Or\'us}},
  \bibinfo {author} {\bibfnamefont {G.}~\bibnamefont {Vidal}}, \bibinfo
  {author} {\bibfnamefont {F.}~\bibnamefont {Verstraete}}, \ and\ \bibinfo
  {author} {\bibfnamefont {J.~I.}\ \bibnamefont {Cirac}},\ }\href {\doibase
  10.1103/PhysRevLett.101.250602} {\bibfield  {journal} {\bibinfo  {journal}
  {Phys. Rev. Lett.}\ }\textbf {\bibinfo {volume} {101}},\ \bibinfo {pages}
  {250602} (\bibinfo {year} {2008})}\BibitemShut {NoStop}%
\bibitem [{\citenamefont {Mart\'{i}n-Delgado}\ \emph
  {et~al.}(2001)\citenamefont {Mart\'{i}n-Delgado}, \citenamefont {Roncaglia},\
  and\ \citenamefont {Sierra}}]{MartinRS2001}%
  \BibitemOpen
  \bibfield  {author} {\bibinfo {author} {\bibfnamefont {M.~A.}\ \bibnamefont
  {Mart\'{i}n-Delgado}}, \bibinfo {author} {\bibfnamefont {M.}~\bibnamefont
  {Roncaglia}}, \ and\ \bibinfo {author} {\bibfnamefont {G.}~\bibnamefont
  {Sierra}},\ }\href {\doibase 10.1103/PhysRevB.64.075117} {\bibfield
  {journal} {\bibinfo  {journal} {Phys. Rev. B}\ }\textbf {\bibinfo {volume}
  {64}},\ \bibinfo {pages} {075117} (\bibinfo {year} {2001})}\BibitemShut
  {NoStop}%
\bibitem [{\citenamefont {Nishino}\ \emph {et~al.}(2001)\citenamefont
  {Nishino}, \citenamefont {Hieida}, \citenamefont {Okunishi}, \citenamefont
  {Maeshima}, \citenamefont {Akutsu},\ and\ \citenamefont
  {Gendiar}}]{NishinoHOMAG2001}%
  \BibitemOpen
  \bibfield  {author} {\bibinfo {author} {\bibfnamefont {T.}~\bibnamefont
  {Nishino}}, \bibinfo {author} {\bibfnamefont {Y.}~\bibnamefont {Hieida}},
  \bibinfo {author} {\bibfnamefont {K.}~\bibnamefont {Okunishi}}, \bibinfo
  {author} {\bibfnamefont {N.}~\bibnamefont {Maeshima}}, \bibinfo {author}
  {\bibfnamefont {Y.}~\bibnamefont {Akutsu}}, \ and\ \bibinfo {author}
  {\bibfnamefont {A.}~\bibnamefont {Gendiar}},\ }\href {\doibase
  10.1143/PTP.105.409} {\bibfield  {journal} {\bibinfo  {journal} {Progress of
  Theoretical Physics}\ }\textbf {\bibinfo {volume} {105}},\ \bibinfo {pages}
  {409} (\bibinfo {year} {2001})}\BibitemShut {NoStop}%
\bibitem [{\citenamefont {Jiang}\ \emph {et~al.}(2008)\citenamefont {Jiang},
  \citenamefont {Weng},\ and\ \citenamefont {Xiang}}]{JiangWX2008}%
  \BibitemOpen
  \bibfield  {author} {\bibinfo {author} {\bibfnamefont {H.~C.}\ \bibnamefont
  {Jiang}}, \bibinfo {author} {\bibfnamefont {Z.~Y.}\ \bibnamefont {Weng}}, \
  and\ \bibinfo {author} {\bibfnamefont {T.}~\bibnamefont {Xiang}},\ }\href
  {\doibase 10.1103/PhysRevLett.101.090603} {\bibfield  {journal} {\bibinfo
  {journal} {Phys. Rev. Lett.}\ }\textbf {\bibinfo {volume} {101}},\ \bibinfo
  {pages} {090603} (\bibinfo {year} {2008})}\BibitemShut {NoStop}%
\bibitem [{\citenamefont {Or\'us}\ and\ \citenamefont
  {Vidal}(2009)}]{OrusV2009}%
  \BibitemOpen
  \bibfield  {author} {\bibinfo {author} {\bibfnamefont {R.}~\bibnamefont
  {Or\'us}}\ and\ \bibinfo {author} {\bibfnamefont {G.}~\bibnamefont {Vidal}},\
  }\href {\doibase 10.1103/PhysRevB.80.094403} {\bibfield  {journal} {\bibinfo
  {journal} {Phys. Rev. B}\ }\textbf {\bibinfo {volume} {80}},\ \bibinfo
  {pages} {094403} (\bibinfo {year} {2009})}\BibitemShut {NoStop}%
\bibitem [{\citenamefont {Baxter}(1968)}]{Baxter1968}%
  \BibitemOpen
  \bibfield  {author} {\bibinfo {author} {\bibfnamefont {R.~J.}\ \bibnamefont
  {Baxter}},\ }\href@noop {} {\bibfield  {journal} {\bibinfo  {journal}
  {Journal of Mathematical Physics}\ }\textbf {\bibinfo {volume} {9}} (\bibinfo
  {year} {1968})}\BibitemShut {NoStop}%
\bibitem [{\citenamefont {Baxter}(1978)}]{Baxter1978}%
  \BibitemOpen
  \bibfield  {author} {\bibinfo {author} {\bibfnamefont {R.~J.}\ \bibnamefont
  {Baxter}},\ }\href {\doibase 10.1007/BF01011693} {\bibfield  {journal}
  {\bibinfo  {journal} {Journal of Statistical Physics}\ }\textbf {\bibinfo
  {volume} {19}},\ \bibinfo {pages} {461} (\bibinfo {year} {1978})}\BibitemShut
  {NoStop}%
\bibitem [{\citenamefont {Baxter}(1982)}]{Baxter_book}%
  \BibitemOpen
  \bibfield  {author} {\bibinfo {author} {\bibfnamefont {R.~J.}\ \bibnamefont
  {Baxter}},\ }\href@noop {} {\emph {\bibinfo {title} {Exactly Solved Models in
  Statistical Mechanics}}}\ (\bibinfo  {publisher} {Academic Press, London},\
  \bibinfo {year} {1982})\BibitemShut {NoStop}%
\bibitem [{\citenamefont {Nishino}\ and\ \citenamefont
  {Okunishi}(1998)}]{NishinoO1998}%
  \BibitemOpen
  \bibfield  {author} {\bibinfo {author} {\bibfnamefont {T.}~\bibnamefont
  {Nishino}}\ and\ \bibinfo {author} {\bibfnamefont {K.}~\bibnamefont
  {Okunishi}},\ }\href {\doibase 10.1143/JPSJ.67.3066} {\bibfield  {journal}
  {\bibinfo  {journal} {Journal of the Physical Society of Japan}\ }\textbf
  {\bibinfo {volume} {67}},\ \bibinfo {pages} {3066} (\bibinfo {year}
  {1998})}\BibitemShut {NoStop}%
\bibitem [{\citenamefont {Corboz}\ \emph {et~al.}(2011)\citenamefont {Corboz},
  \citenamefont {White}, \citenamefont {Vidal},\ and\ \citenamefont
  {Troyer}}]{CorbozWV2011}%
  \BibitemOpen
  \bibfield  {author} {\bibinfo {author} {\bibfnamefont {P.}~\bibnamefont
  {Corboz}}, \bibinfo {author} {\bibfnamefont {S.~R.}\ \bibnamefont {White}},
  \bibinfo {author} {\bibfnamefont {G.}~\bibnamefont {Vidal}}, \ and\ \bibinfo
  {author} {\bibfnamefont {M.}~\bibnamefont {Troyer}},\ }\href {\doibase
  10.1103/PhysRevB.84.041108} {\bibfield  {journal} {\bibinfo  {journal} {Phys.
  Rev. B}\ }\textbf {\bibinfo {volume} {84}},\ \bibinfo {pages} {041108}
  (\bibinfo {year} {2011})}\BibitemShut {NoStop}%
\bibitem [{\citenamefont {Corboz}\ \emph {et~al.}(2014)\citenamefont {Corboz},
  \citenamefont {Rice},\ and\ \citenamefont {Troyer}}]{CorbozRT2014}%
  \BibitemOpen
  \bibfield  {author} {\bibinfo {author} {\bibfnamefont {P.}~\bibnamefont
  {Corboz}}, \bibinfo {author} {\bibfnamefont {T.~M.}\ \bibnamefont {Rice}}, \
  and\ \bibinfo {author} {\bibfnamefont {M.}~\bibnamefont {Troyer}},\ }\href
  {\doibase 10.1103/PhysRevLett.113.046402} {\bibfield  {journal} {\bibinfo
  {journal} {Phys. Rev. Lett.}\ }\textbf {\bibinfo {volume} {113}},\ \bibinfo
  {pages} {046402} (\bibinfo {year} {2014})}\BibitemShut {NoStop}%
\bibitem [{\citenamefont {Jos\'e}\ \emph {et~al.}(1977)\citenamefont {Jos\'e},
  \citenamefont {Kadanoff}, \citenamefont {Kirkpatrick},\ and\ \citenamefont
  {Nelson}}]{JoseKKN1977}%
  \BibitemOpen
  \bibfield  {author} {\bibinfo {author} {\bibfnamefont {J.~V.}\ \bibnamefont
  {Jos\'e}}, \bibinfo {author} {\bibfnamefont {L.~P.}\ \bibnamefont
  {Kadanoff}}, \bibinfo {author} {\bibfnamefont {S.}~\bibnamefont
  {Kirkpatrick}}, \ and\ \bibinfo {author} {\bibfnamefont {D.~R.}\ \bibnamefont
  {Nelson}},\ }\href {\doibase 10.1103/PhysRevB.16.1217} {\bibfield  {journal}
  {\bibinfo  {journal} {Phys. Rev. B}\ }\textbf {\bibinfo {volume} {16}},\
  \bibinfo {pages} {1217} (\bibinfo {year} {1977})}\BibitemShut {NoStop}%
\bibitem [{Note1()}]{Note1}%
  \BibitemOpen
  \bibinfo {note} {$\protect \mathcal {H}\Phi $ is available at
  https://github.com/QLMS/HPhi}\BibitemShut {NoStop}%
\bibitem [{\citenamefont {Suzuki}(1976)}]{Suzuki1976}%
  \BibitemOpen
  \bibfield  {author} {\bibinfo {author} {\bibfnamefont {M.}~\bibnamefont
  {Suzuki}},\ }\href {\doibase 10.1143/PTP.56.1454} {\bibfield  {journal}
  {\bibinfo  {journal} {Progress of Theoretical Physics}\ }\textbf {\bibinfo
  {volume} {56}},\ \bibinfo {pages} {1454} (\bibinfo {year}
  {1976})}\BibitemShut {NoStop}%
\bibitem [{\citenamefont {Trotter}(1959)}]{Trotter1959}%
  \BibitemOpen
  \bibfield  {author} {\bibinfo {author} {\bibfnamefont {H.~F.}\ \bibnamefont
  {Trotter}},\ }\href {\doibase 10.1090/S0002-9939-1959-0108732-6} {\bibfield
  {journal} {\bibinfo  {journal} {Proceedings of the American Mathematical
  Society}\ }\textbf {\bibinfo {volume} {10}},\ \bibinfo {pages} {545}
  (\bibinfo {year} {1959})}\BibitemShut {NoStop}%
\bibitem [{\citenamefont {Corboz}\ \emph {et~al.}(2010)\citenamefont {Corboz},
  \citenamefont {Jordan},\ and\ \citenamefont {Vidal}}]{CorbozJV2010}%
  \BibitemOpen
  \bibfield  {author} {\bibinfo {author} {\bibfnamefont {P.}~\bibnamefont
  {Corboz}}, \bibinfo {author} {\bibfnamefont {J.}~\bibnamefont {Jordan}}, \
  and\ \bibinfo {author} {\bibfnamefont {G.}~\bibnamefont {Vidal}},\ }\href
  {\doibase 10.1103/PhysRevB.82.245119} {\bibfield  {journal} {\bibinfo
  {journal} {Phys. Rev. B}\ }\textbf {\bibinfo {volume} {82}},\ \bibinfo
  {pages} {245119} (\bibinfo {year} {2010})}\BibitemShut {NoStop}%
\bibitem [{\citenamefont {Wang}\ \emph {et~al.}(2011)\citenamefont {Wang},
  \citenamefont {Pi\ifmmode~\check{z}\else \v{z}\fi{}orn},\ and\ \citenamefont
  {Verstraete}}]{WangPV2011}%
  \BibitemOpen
  \bibfield  {author} {\bibinfo {author} {\bibfnamefont {L.}~\bibnamefont
  {Wang}}, \bibinfo {author} {\bibfnamefont {I.}~\bibnamefont
  {Pi\ifmmode~\check{z}\else \v{z}\fi{}orn}}, \ and\ \bibinfo {author}
  {\bibfnamefont {F.}~\bibnamefont {Verstraete}},\ }\href {\doibase
  10.1103/PhysRevB.83.134421} {\bibfield  {journal} {\bibinfo  {journal} {Phys.
  Rev. B}\ }\textbf {\bibinfo {volume} {83}},\ \bibinfo {pages} {134421}
  (\bibinfo {year} {2011})}\BibitemShut {NoStop}%
\bibitem [{\citenamefont {Depenbrock}\ and\ \citenamefont
  {Pollmann}(2013)}]{DepenbrockP2013}%
  \BibitemOpen
  \bibfield  {author} {\bibinfo {author} {\bibfnamefont {S.}~\bibnamefont
  {Depenbrock}}\ and\ \bibinfo {author} {\bibfnamefont {F.}~\bibnamefont
  {Pollmann}},\ }\href {\doibase 10.1103/PhysRevB.88.035138} {\bibfield
  {journal} {\bibinfo  {journal} {Phys. Rev. B}\ }\textbf {\bibinfo {volume}
  {88}},\ \bibinfo {pages} {035138} (\bibinfo {year} {2013})}\BibitemShut
  {NoStop}%
\end{thebibliography}%

\end{document}